\documentclass[12pt]{article}
\pdfoutput=1
\usepackage{epsf,amsfonts,amssymb,epsfig,amsmath,mathtools,graphics,slashed}
\usepackage{hyperref,hep,graphicx,subfig}
\usepackage{rotating}

\newcommand{\nn}{\notag \\}

\newcommand{\dd}{\mathrm{d}}

\makeatletter

\@addtoreset{equation}{section}
\makeatother

\begin{document}

\begin{titlepage}

\vfill

\begin{flushright}
Imperial/TP/2017/JG/05\\
DCPT-17/41
\end{flushright}

\vfill

\begin{center}
   \baselineskip=16pt
   {\Large\bf Boomerang RG flows with\\intermediate conformal invariance}
  \vskip 1.5cm
  \vskip 1.5cm
Aristomenis Donos$^1$, Jerome P. Gauntlett$^2$\\ Christopher Rosen$^2$ and Omar Sosa-Rodriguez$^1$\\
     \vskip .6cm
     \begin{small}
      \textit{$^1$Centre for Particle Theory and Department of Mathematical Sciences\\Durham University, Durham, DH1 3LE, U.K.}
        \end{small}\\    
         \begin{small}\vskip .6cm
      \textit{$^2$Blackett Laboratory, 
  Imperial College\\ London, SW7 2AZ, U.K.}
        \end{small}\\
        \end{center}
\begin{center}
\textbf{Abstract}
\end{center}
\begin{quote}
For a class of $D=5$ holographic models we construct boomerang RG flow solutions that start in the UV at an $AdS_5$ vacuum
and end up at the same vacuum in the IR. The RG flows are driven by deformations by
relevant operators that explicitly break translation invariance. For specific models, such that they admit another $AdS_5$
solution, $AdS_5^c$,  
we show that for large enough deformations the RG flows approach an intermediate scaling regime with approximate conformal invariance
governed by $AdS^c_5$. For these flows we calculate the holographic entanglement entropy and the entropic $c$-function for the RG flows. The latter is not monotonic, but it does encapsulate the degrees of freedom in each scaling region. 
For a different set of models, we find boomerang RG flows with intermediate scaling
governed by an $AdS_2\times\mathbb{R}^3$ solution which breaks translation invariance. Furthermore, for large enough deformations
these models have interesting and novel
thermal insulating ground states for which the entropy vanishes as the temperature goes to zero, but not as a power-law. 
Remarkably, the thermal diffusivity and the butterfly velocity for these new insulating ground states are related via $D=Ev^2_B/(2\pi T)$, with
$E(T)\to 0.5$ as $T\to 0$.
\end{quote}

\vfill

\end{titlepage}

\setcounter{equation}{0}
\section{Introduction}

A boomerang RG flow starts at an RG fixed point in the UV 
and then flows to exactly the same RG fixed point in the IR \cite{Donos:2017ljs}. A particularly interesting realisation is when the RG
fixed point is conformally invariant. In this context, in order to be consistent with either the letter or the
spirit of $c$-theorems, the deformations of the UV fixed point which are driving the RG flow should necessarily break Poincar\'e invariance. 

Various examples of such boomerang RG flows have been explicitly constructed within the context of the AdS/CFT correspondence
\cite{Chesler:2013qla,Donos:2014gya,Donos:2016zpf,Donos:2017ljs}. 
For example, within pure $D=5$ Einstein gravity with a negative cosmological constant, 
and hence of relevance to the most general class of $d=4$ CFTs with a holographic dual, boomerang RG flows associated with
helical deformations
of the metric were constructed in \cite{Donos:2014gya}.
By contrast the boomerang flows of \cite{Chesler:2013qla} are associated with CFTs with a global $U(1)$ symmetry and with a spatially varying chemical potential. In another direction, the constructions in \cite{Donos:2016zpf,Donos:2017ljs} involve deformations of operators dual to bulk scalar fields. All of these examples describe CFTs which have been deformed by operators which explicitly break translation invariance in one or more of the spatial directions and hence are special examples of holographic lattices \cite{Horowitz:2012ky}. One motivation for studying holographic lattices is that they provide a natural framework for studying thermal and electric transport with finite DC conductivities.

The examples studied in \cite{Chesler:2013qla,Donos:2014gya,Donos:2016zpf,Donos:2017ljs} all involve deformations with a single spatial Fourier mode and,
for small enough deformations, a perturbative expansion can be used to argue for the existence of 
boomerang flows. Indeed, the perturbative deformation of the bulk field that is dual to the deforming operator exponentially dies out near the Poincar\'e horizon
and hence is not expected\footnote{A subtlety is that one needs to check that the expansion does not generate constant Fourier modes
which can change the IR.} 
to modify the IR. 
An interesting feature of the specific top-down examples constructed in \cite{Donos:2014gya,Donos:2016zpf,Donos:2017ljs}
is that the boomerang flows actually persist for arbitrarily large deformations, which {\it a priori}, is not guaranteed.
Furthermore, it is particularly
interesting that for sufficiently large deformations the boomerang flows \cite{Donos:2014gya,Donos:2016zpf,Donos:2017ljs}
exhibit one or more intermediate scaling regimes, where the solution approaches, somewhere in the bulk,
a configuration with scaling properties.
In the constructions of \cite{Donos:2016zpf}, which involved deformations
of the axion and dilaton in the context of $AdS_5\times X_5$ solutions of type IIB supergravity, the intermediate scaling is dominated by a fixed 
point solution with Lifshitz-like scaling \cite{Azeyanagi:2009pr}. By contrast, the constructions in \cite{Donos:2017ljs} were made in the context of $D=11$ supergravity and are of relevance to ABJM theory. In these examples, for large enough deformations, the boomerang flows approach two intermediate scaling regimes in succession, each associated with hyperscaling violation. 

The original aim of this paper was to construct boomerang RG flows in $D=5$ which have an intermediate scaling regime 
governed by another $AdS_5$ factor associated with approximate $d=4$ conformal invariance.
We have not yet been able to find top-down examples but, as we will see, it is quite straightforward to construct bottom-up examples. 
As in \cite{Donos:2016zpf,Donos:2017ljs}, we will utilise a Q-lattice construction \cite{Donos:2013eha} in which we exploit a global symmetry of the bulk spacetime
in order to develop an ansatz for the bulk fields in which the dependence on the spatial directions of the CFT is solved exactly. This leads
to a system of ordinary differential equations for a set of functions that just depend on the holographic radial coordinate which are 
then amenable to straightforward numerical integration. 

A key ingredient in our construction is to have a bulk theory that admits a Poincar\'e invariant domain wall solution that
flows between $AdS_5^0$ in the UV and another $AdS_5^c$ in the IR. We demand that this domain wall flow 
is driven by deformations of relevant operators in the UV CFT, with scaling dimension $\Delta$, 
and hence is parametrised by a dimensionful parameter $\Gamma$.
By conformal invariance all values of $\Gamma$ are physically equivalent for these Poincar\'e invariant RG flows.
Within a Q-lattice ansatz, we then consider deformations by the same relevant operators which also have a dependence on the spatial directions of the CFT, parametrised by a wave number $k$. This gives rise to a one parameter family of associated RG flows, parametrised by a dimensionless number $\Gamma/k^{4-\Delta}$. For small values of $\Gamma/k^{4-\Delta}$ we can easily show that
we must have boomerang RG flows using a perturbative construction. For larger values of $\Gamma/k^{4-\Delta}$ the existence of the boomerang
flows must be established numerically. When they do exist, though, since large values of $\Gamma/k^{4-\Delta}$ can be achieved by
holding $\Gamma$ fixed and taking $k\to 0$, one can expect that the boomerang RG flows should start to track the Poincar\'e invariant flow
and hence exhibit an intermediate scaling regime with conformal invariance that is governed by the $AdS_5^c$ fixed point solution. 

For holographic RG flows with intermediate scaling regimes, which have also been extensively studied in other contexts (e.g. \cite{Harrison:2012vy,Bao:2012yt,Bhattacharya:2012zu,Jain:2014vka,Kundu:2012jn,Donos:2012yi}), it is
of interest to investigate to what extent the scaling regime imprints itself on the scaling behaviour of physical observables. For example, one 
might expect that the spectral weight of operators as a function of frequency, $\omega$, should exhibit scaling for a range of $\omega$
dictated by the range of the radial region of the RG flow which has intermediate scaling. This issue was discussed in \cite{Donos:2017ljs} using matching arguments (for a related discussion see \cite{Bhattacharya:2014dea}). It was shown that while intermediate scaling behaviour is not guaranteed it will manifest itself providing sufficient conditions on the effective potential for the bulk fluctuations about the RG flow solutions are met \cite{Donos:2017ljs}.
In this paper we make a complementary discussion by examining how the holographic entanglement entropy behaves for the new boomerang RG flows. In particular, by calculating the entanglement entropy of a strip geometry of width $l$, we analyse the
behaviour of the entropic `$c$-function', ${C}(l)$ \cite{2007JPhA...40.7031C,Nishioka:2009un,Myers:2012ed} (see also \cite{Liu:2012eea}). While ${C}(l)$ is not monotonic along the boomerang flow, as it is for Poincar\'e invariant RG flows, it does
effectively encapsulate the correct scaling of the degrees of freedom of CFT in the UV and IR as well as the CFT in the
intermediate scaling regime.

We will study a class of $D=5$ models with a quartic potential for the scalar fields that depend on two real parameters.
The constructions summarised above are for certain values of the parameters, such that the models admit both the $AdS_5^0$ vacuum and also
the $AdS_5^c$ solution (in fact there will be two $AdS_5^c$ related by a $\mathbb{Z}_2$ symmetry). Interestingly, for different values of the parameters there
is no longer an $AdS_5^c$ solution but there is an $AdS_2\times\mathbb{R}^3$ solution which breaks translations in all of the spatial directions. 

\begin{figure}[h!]
\centering
\includegraphics[scale=0.35]{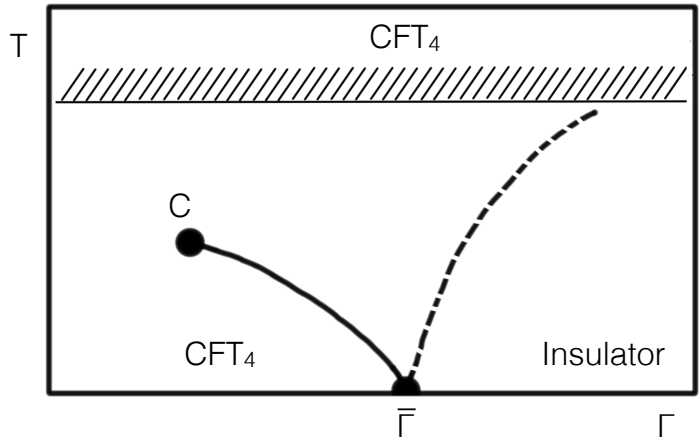}
\caption{Schematic phase diagram as a function of the deformation parameter
$\Gamma$ for models with $m^2 =-15/4$, $\xi = -1/4$ discussed in section \ref{qc}. The $T=0$ ground states are RG flows from $AdS_5$ in the UV, dual to some CFT$_4$, to various behaviours in
the IR: for $\Gamma<\bar\Gamma$ we have $AdS_5$ (boomerang RG flows), for
$\Gamma=\bar\Gamma$ we have $AdS_2\times\mathbb{R}^3$ (locally quantum critical ground states),
and for $\Gamma>\bar\Gamma$ we have singular thermal insulating behaviour. There is line of first order phase transitions that end in the critical point $C$. Intermediate scaling governed by the $AdS_2\times\mathbb{R}^3$ solution is present in the quantum critical wedge bounded by the first order line and the dashed line. For high temperatures the scaling is fixed by the $AdS_5$ solution in the UV.} \label{schempt}
\end{figure}

In the second part of the paper, starting in section \ref{qc}, we will investigate models with boomerang flows that have intermediate scaling
governed by such locally quantum critical $AdS_2\times\mathbb{R}^3$ solutions. While there are some similarities to the previous constructions there 
are also some interesting differences. The RG flows from $AdS_5^0$ in the UV to $AdS_2\times\mathbb{R}^3$ in the IR
now exist for a specific value of the dimensionless deformation parameter $\Gamma/k^{4-\Delta}\equiv \bar\Gamma$. 
Focussing on a specific model, we find that the boomerang RG flows only exist in the range
$0\le \Gamma/k^{4-\Delta}\le\bar\Gamma$, and moreover, have increasingly large intermediate scaling behaviour determined by 
the $AdS_2\times\mathbb{R}^3$ solution as $\Gamma/k^{4-\Delta}\to \bar\Gamma$. In order to understand the RG flows for 
$\Gamma/k^{4-\Delta}>\bar\Gamma$ we construct finite temperature black holes and then cool them down to very low temperatures.
This investigation reveals an interesting phase diagram schematically presented in figure \ref{schempt}. For a range of 
$\Gamma/k^{4-\Delta}\le\bar\Gamma$ there is a line of first order phase transitions ending on the $AdS_2\times\mathbb{R}^3$ fixed point
at $T=0$ and on a finite temperature critical point. Furthermore, the $T=0$ ground states for $\Gamma/k^{4-\Delta}>\bar\Gamma$
are singular\footnote{Since these ground states are obtained by cooling down black hole solutions, they are necessarily
``good singularities" in the sense of \cite{Gubser:2000nd}.} and, by calculating the behaviour of the thermal DC conductivity, $\kappa$, as a function of temperature,
we conclude that they are thermally insulating ground states.

The behaviour of the entropy as a function of temperature is not a power law for the new insulating ground states,
in contrast to those constructed in \cite{Donos:2012js,Donos:2014uba,Gouteraux:2014hca,Donos:2014oha}. 
This in itself makes them worthy of further study. One additional calculation that we carry out here is 
motivated by the various investigations aiming to elucidate universal connections between diffusion and quantum chaos 
in holography \cite{Hartnoll:2014lpa,Blake:2016wvh,Blake:2016sud} (see also \cite{Lucas:2016yfl,Ling:2016ibq,Blake:2016jnn,Davison:2016ngz,Baggioli:2016pia,Wu:2017mdl,Kim:2017dgz,Baggioli:2017ojd,Blake:2017qgd,Wu:2017exh,Giataganas:2017koz, Grozdanov:2017ajz, Lucas:2017ibu}). We determine the thermal diffusion constant, $D$, using the Einstein relation $D\equiv \kappa/c$, where $c$ is the specific
heat. We also calculate the butterfly velocity, $v_B$, by analysing a shockwave on the black hole solution as in
\cite{Shenker:2013pqa,Maldacena:2015waa} (see also \cite{Sfetsos:1994xa}).
Remarkably, we find 
\begin{align}
D=E\frac{v^2_B}{2\pi T}\,,
\end{align}
with the dimensionless quantity $E(T)\to 0.5$ as $T\to 0$. This is the first example of such a relationship for ground states
without power law behaviour.  We have also made some other constructions for models with 
slightly different parameters\footnote{For certain models we also find a novel non-uniqueness of the boomerang RG flows which we discuss in appendix \ref{othervals}.} to those in figure \ref{schempt}, 
again finding insulating ground states with $E$ unchanged.

\section{General set up}\label{gensetup}
Consider an action in $D=5$ spacetime dimensions of the form
\begin{equation}\label{eq:act}
S =\frac{1}{16 \pi G}\int\dd^5x\,\sqrt{-g}\Big(R+12+\mathcal{L}_z \Big)\,,
\end{equation}
where $\mathcal{L}_z$ describes a sigma model for three complex scalars $z^\alpha$.
In order to construct the Q-lattice solutions of interest, we will take $\mathcal{L}_z$ to have a $U(1)^3$ global symmetry
and consider
\begin{equation}\label{eq:lag}
\mathcal{L}_z = \sum_\alpha\left(-\frac{1}{2}\,\partial_\mu z^{\alpha}\partial^\mu \bar{z}^{\bar{\alpha}}
-\frac{1}{2}m^2\,z^{\alpha}\bar{z}^{\bar{\alpha}}-\frac{1}{3}\xi\,\left(z^{\alpha}\bar{z}^{\bar{\alpha}}\right)^2\right)\,,
\end{equation}
where $m^2,\xi$ are two free parameters. The equations of motion admit a unit radius $AdS_5^0$ vacuum solution with $z^\alpha=0$ 
which is dual to a CFT in $d=4$. 
In these units $1/16 \pi G$ is a measure of the number of degrees of freedom in the dual CFT, scaling
like $N^2$, at large $N$. In a slight abuse of notation we will set factors of $16 \pi G$ to unity in the following since this simplifies some formulae and
the factors can easily be reinstated in physical quantities as needed.

We are interested in studying specific isotropic deformations of this CFT that break translations in three spatial directions.
To do this we exploit the $U(1)^3$ global symmetry and consider the $Q$-lattice ansatz 
\begin{align}\label{eq:Qans}
\dd s^2 &= -g(r)e^{-\chi(r)}\dd t^2+\frac{\dd r^2}{g(r)}+r^2dx^\alpha dx^\alpha
 \,,\nn
z^{\alpha} &= \gamma(r)e^{i k\, x^\alpha}\,,
\end{align}
where $x^\alpha \in \{x,y,z\}$ are the spatial directions of the field theory. Notice that a simultaneous translation and a $U(1)^3$ transformation
preserves this ansatz. 
The associated equations of motion are given by
\begin{align} \label{eq:chiEQ}
0 & =\, \chi' + r\,\gamma'^2\,,\nn
0 & = \, g'+g \left(\frac{1}{2}
   r \gamma '^2+\frac{2}{r}\right)+\gamma^2\frac {\left(k^2+m^2 r^2\right)}{2
   r}+\frac{1}{3} \xi  r \gamma^4-4 r\,,\nn
 0 & = \, \gamma
   ''+\gamma ' \left(\frac{g'}{g}-\frac{\chi
   '}{2}+\frac{3}{r}\right)-\gamma \frac{\left(m^2 r^2+k^2\right)}{r^2 g}-\frac{4 \xi  \gamma^3}{3 g}\,.
\end{align}

In sections\footnote{In section \ref{qc} we will consider boomerang flows with an intermediate $AdS_2\times\mathbb{R}^3$ solution and it is convenient to use a different radial coordinate to that of \eqref{eq:Qans}.}  \ref{gensetup}-\ref{ententropy}, we will be focussing on boomerang flows from $AdS_5^0$ in the UV to $AdS_5^0$ in the IR that have an intermediate 
scaling behaviour governed by a different $AdS_5$ solution.
We will choose the parameters $m^2,\xi$ so that there are, in fact, three stable $AdS_5$ solutions 
with constant $\gamma$ (and all with $k=0$). Writing the $AdS_5$ metric in Poincar\'e coordinates as
\begin{equation}\label{eq:AdS}
\dd s^2 = -r^2\dd t^2 +r^2 \dd\vec{x}^2+\frac{L^2}{r^2}\dd r^2\,,
\end{equation}
the UV $AdS_5$ vacuum solution, which we call $AdS_5^0$, has 
\begin{align}
L_0^2 = 1 \qquad \mathrm{and}\qquad \gamma_0= 0\,,
\end{align}
while the other two $AdS_5$ solutions, which we call $AdS_5^c$, have
\begin{align}\label{irads}
L_c^2 = \frac{64\xi}{3 m^4+64\xi} \qquad \mathrm{and}\qquad \gamma_c =  \pm \sqrt{\frac{-3m^2}{4\xi}}\,.
\end{align}
In order to have suitable relevant and irrelevant scalar operators in the UV and IR, with conformal dimensions
$\Delta_0$ and $\Delta_c$, respectively, we demand that
\begin{align}\label{real}
\Delta_0\equiv 2+\sqrt{4+m^2}< 4\,,\qquad
\Delta_c \equiv   2+\sqrt{4-2m^2L^2_c}>4\,.
\end{align}
In sections \ref{gensetup}-\ref{ententropy} we will focus the specific values of $m,\xi$ given
by 
\begin{align}\label{mxfirst}
m^2 =-15/4\,,\qquad \xi = 675/512\,,
\end{align}
corresponding to having a relevant scalar operator with dimension\footnote{Note that, with this value of $m^2$ we can also, if we wish, do an alternative quantisation of the scalar field leading to $\Delta=3/2$ \cite{Klebanov:1999tb}.}  
$\Delta_0=5/2$ in the CFT dual to the $AdS_5^0$ UV vacuum 
and an irrelevant scalar operator with dimension $\Delta_c=5$ in the CFT dual to $AdS_5^c$. 
Furthermore, for the $AdS_5^c$ vacuum we have $L^2_c=2/3$ and
$\gamma_c=\pm (32/15)^{1/2}$.  

\subsection{Poincar\'e invariant domain wall flows: $AdS_5^0\to AdS_5^c$} 
With the set up just described there are standard Poincar\'e invariant domain wall solutions, with $k=0$ in \eqref{eq:Qans}, 
that approach the unit radius $AdS_5^0$ vacuum solution in
the UV and then approach one of the two $AdS_5^c$ solutions \eqref{irads} in the IR. As there is a $\mathbb{Z}_2$ symmetry relating
these IR vacua, without loss of generality we can focus on the solution with positive $\gamma_c$.

In the UV, as $r\to\infty$, the solutions have a radial expansion of the form
\begin{align}\label{uvcons}
g = {r^2}+\dots \qquad \chi = \chi_{UV} + \dots  \qquad \gamma = {\Gamma}r^{\Delta_0 -4}+\dots\,,
\end{align}
and we set $\chi_{UV}=0$. 
In the IR, as $r\to 0$, we have the expansion
\begin{align}
g = \frac{r^2}{L_c^2}+\dots \qquad \chi = \chi_0 + \dots  \qquad \gamma = \gamma_c + f_0 r^{\Delta_c -4}+\dots.
\end{align}
The boundary conditions \eqref{uvcons} are associated with deformations of the
CFT dual to $AdS_5^0$,  parametrised by $\Gamma$, using the real parts of the three operators $\mathcal{O}_\alpha$ that are dual to the three complex scalars $z^\alpha$. Note that within the ansatz \eqref{eq:Qans}, with $k=0$, the deformation of all three operators are the same. There are additional domain wall solutions flowing to the same $AdS_5^c$ solution in the IR that lie outside this ansatz, but they will not play a role in the sequel.
We also note that by conformal invariance, the $k=0$ domain walls with different values of $\Gamma$ are all physically equivalent.

We have explicitly constructed these domain wall solutions, for the specific values of $m^2,\xi$ given in \eqref{mxfirst},
using numerical shooting techniques. We did this both by shooting from the UV and the IR and then matching at an intermediate point, as well as shooting out just from the IR\footnote{In this approach the solutions generically reach the UV with non-vanishing constant parameter $\chi\to \chi_{UV}$, as $r\to\infty$,
so a simple rescaling of the time coordinate is necessary to bring the asymptotic metric to the canonical form (\ref{eq:AdS}).}, with excellent numerical agreement.

\section{Boomerang RG Flows}
We now want to consider RG flows, with $k\ne 0$ in the ansatz \eqref{eq:Qans}, that are seeded by deformations of the UV CFT
by relevant operators 
which break translation invariance. Imposing the boundary conditions \eqref{uvcons} in the UV now corresponds to
deformations of the real and imaginary parts of the operators $\mathcal{O}_\alpha$ having spatial 
modulation of the form $\cos kx$ and $\sin kx$, respectively. For the specific values of $m,\xi$ given earlier, with $\Delta_0=5/2$,
we deduce that there is a one-parameter family of RG flows that are parametrised by the dimensionless number $\Gamma/k^{3/2}$.

\subsection{Perturbative Analysis}
For $\Gamma/k^{3/2}\ll1$, it is straightforward to argue that the RG flows must be boomerang flows, returning to the same $AdS_5^0$
vacuum (with $\gamma=0$) in the IR. Indeed we can construct the RG flows in a perturbative expansion about the $AdS_5^0$ 
vacuum solution. Starting with the linearised scalar equation of motion in the $AdS_5^0$ vacuum, we find that
the leading order solution that satisfies the UV boundary conditions
and is regular at $r=0$, is of the form
\begin{align}\label{eq:expg1}
\gamma =&\, \frac{k^{3/2}}{r^{3/2}}e^{-k/r}\left(\frac{\Gamma}{k^{3/2}}\right)+\ldots\,.
\end{align}
This solution back reacts on the metric at order $(\Gamma/k^{3/2})^2$ and the explicit form of the metric functions, 
satisfying the correct UV boundary conditions, are given at this order by
\begin{align}\label{eq:expg2}
g=&\, r^2\left[ 1- \frac{k^3}{r^3}e^{-2k/r}\frac{1}{4}\Big(-3+2\frac{k}{r}\Big)
\left(\frac{\Gamma}{k^{3/2}}\right)^2+\ldots\right]\,,\nn
\chi = &\, \frac{3}{16}\left(\frac{\Gamma}{k^{3/2}}\right)^2-e^{-2k/r}\frac{1}{16}\Big(3+6 \frac{k}{r} + 6 \frac{k^2}{r^2} - 8 \frac{k^3}{r^3}
+8 \frac{k^4}{r^4}  \Big)\left(\frac{\Gamma}{k^{3/2}}\right)^2+\ldots\,.
\end{align}

It is clear that in the IR as $r\to 0$, the metric exponentially approaches exactly the same $AdS_5^0$ solution as the UV, with the scale of approach set by $k$. The only difference is that there is a renormalisation of length scales, which is captured by the `index of refraction' $n$ 
\cite{Gubser:2009gp} defined by 
\begin{equation}\label{refinddef}
n \equiv 
e^{\frac{1}{2}(\chi_{IR}-\chi_{UV})}\,.
\end{equation}
At leading order in the expansion we immediately deduce that
\begin{align}\label{pertrefind}
n=1+\frac{3}{32}\left(\frac{\Gamma}{k^{3/2}}\right)^2+\ldots\,.
\end{align}
That the index of refraction is bigger than one is an example of a more general result. Indeed
returning to the equations of motion \eqref{eq:chiEQ} we immediately deduce that $\chi' \le 0$ and hence
$\chi_{IR}\ge\chi_{UV}$, with the
equality realised only for flows in which the scalar does not run.

\subsection{Numerical Boomerang Flows}\label{numbf}
\begin{figure}
\centering
\includegraphics[scale=0.4]{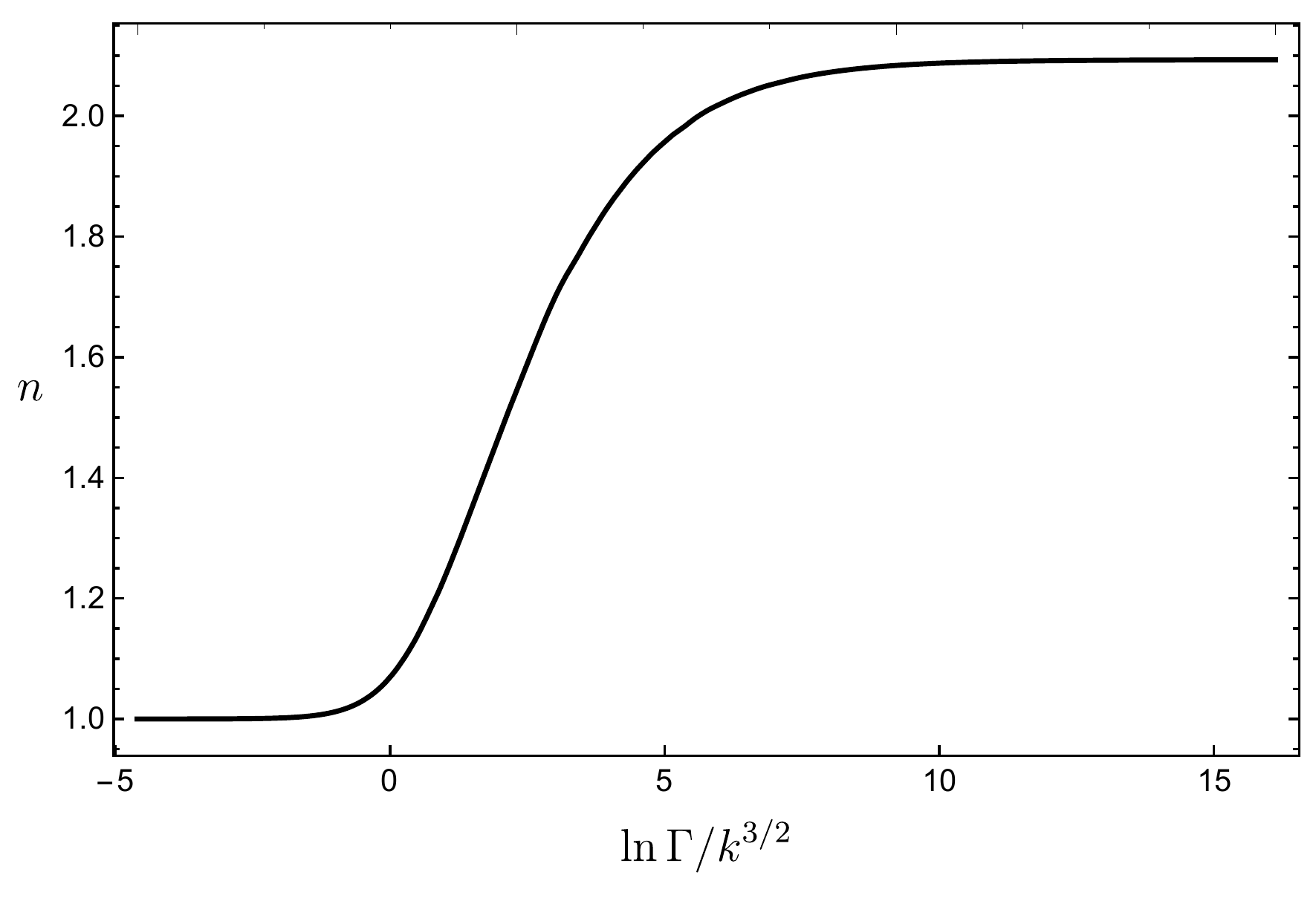}
\caption{Plot of the refractive index $n$, defined in \eqref{refinddef}, as a function of the dimensionless deformation parameter
$\Gamma/k^{3/2}$, for boomerang RG flows with $m^2 =-15/4$ and $\xi = 675/512$.} \label{refind}
\end{figure}

To determine what happens for larger values of $\Gamma/k^{3/2}$ we need to construct the solutions numerically. 
In the IR we develop the expansion
\begin{align}\label{eq:expg3}
g=&\, r^2\left[ 1- \frac{k^3}{r^3}e^{-2k/r}\frac{1}{4}\Big(-3+2\frac{k}{r}\Big)
\left(\frac{C_\gamma}{k^{3/2}}\right)^2+\ldots\right]\,,\nn
\chi = &\, \chi_0-e^{-2k/r}\frac{1}{16}\Big(3+6 \frac{k}{r} + 6 \frac{k^2}{r^2} - 8 \frac{k^3}{r^3}
+8 \frac{k^4}{r^4}  \Big)\left(\frac{C_\gamma}{k^{3/2}}\right)^2+\ldots\,,\nn
\gamma =&\, \frac{k^{3/2}}{r^{3/2}}e^{-k/r}\left(\frac{C_\gamma}{k^{3/2}}\right)+\ldots\,,
\end{align}
where $C_\gamma,\chi_0$ are constants, and demand that the solutions match onto the UV boundary conditions \eqref{uvcons}.
In the range $0<\Gamma/k^{3/2}<10^7$ we find that the RG flows are always boomerang flows, returning to the same $AdS_5^0$ in the IR. Furthermore, we have no reason to suspect that this behaviour will not persist for larger values of $\Gamma/k^{3/2}$.
 In figure \ref{refind} we have presented the index of refraction $n$ for the flows. For small values of $\Gamma/k^{3/2}$ we recover the behaviour \eqref{pertrefind}, as expected. For very large $\Gamma/k^{3/2}$
we find that $n$ appears\footnote{This can be contrasted with the $D=4$boomerang flows constructed in \cite{Donos:2017ljs}, where it was unbounded.} to asymptote to a constant, with $n\sim 2.09$. 

\begin{figure}[h!]
\centering
\includegraphics[scale=0.4]{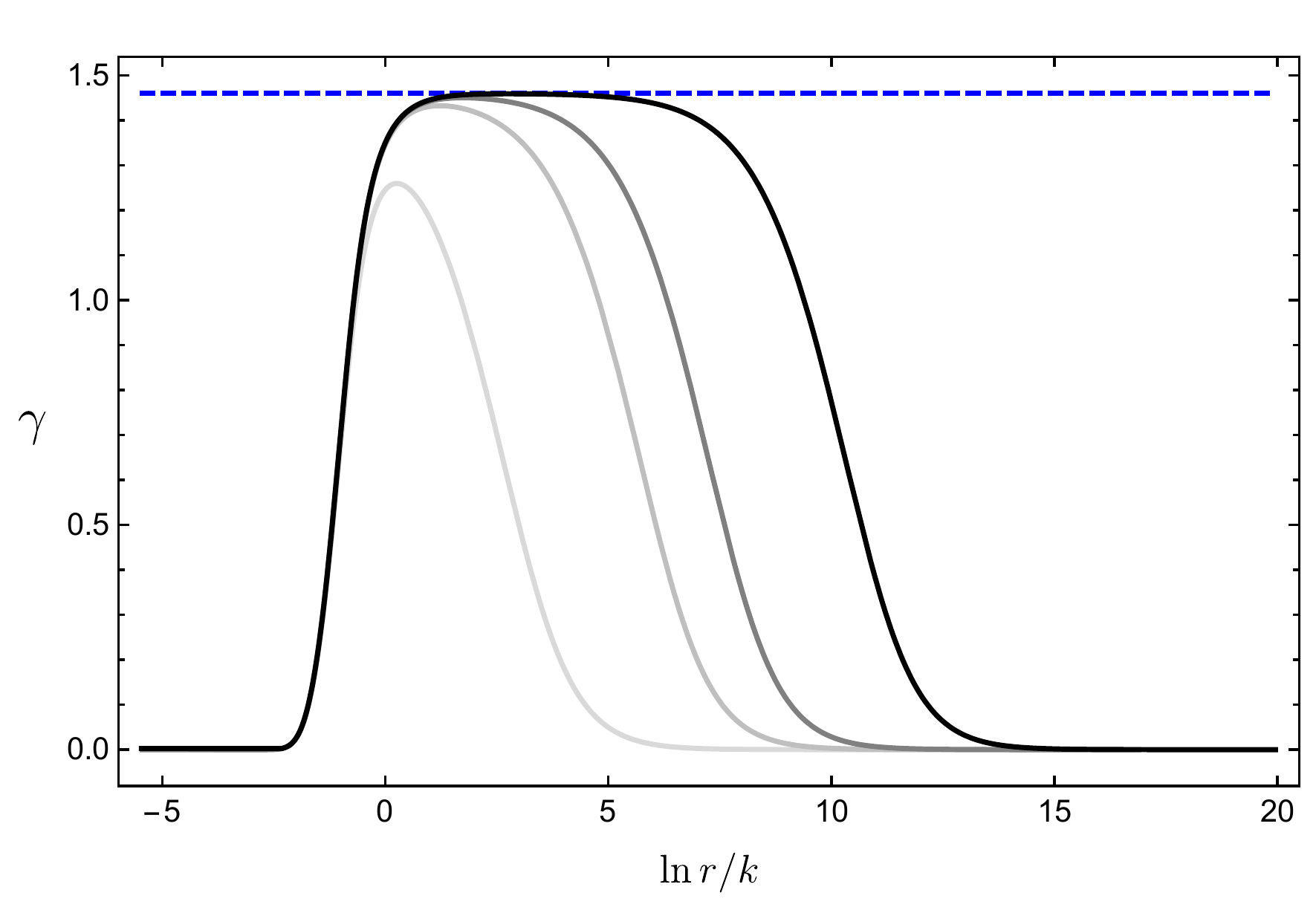}\quad
\includegraphics[scale=0.4]{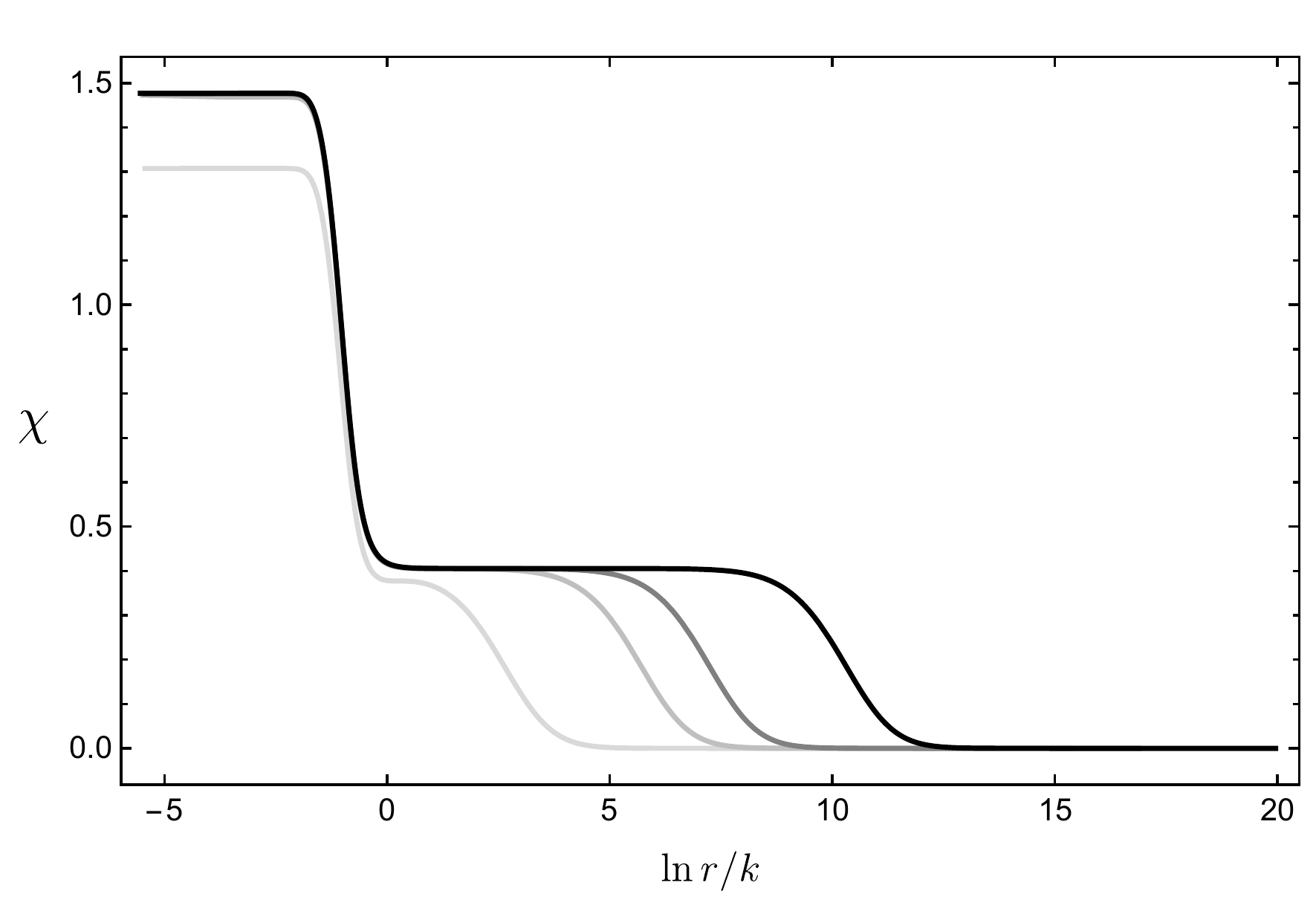}\\
\includegraphics[scale=0.5]{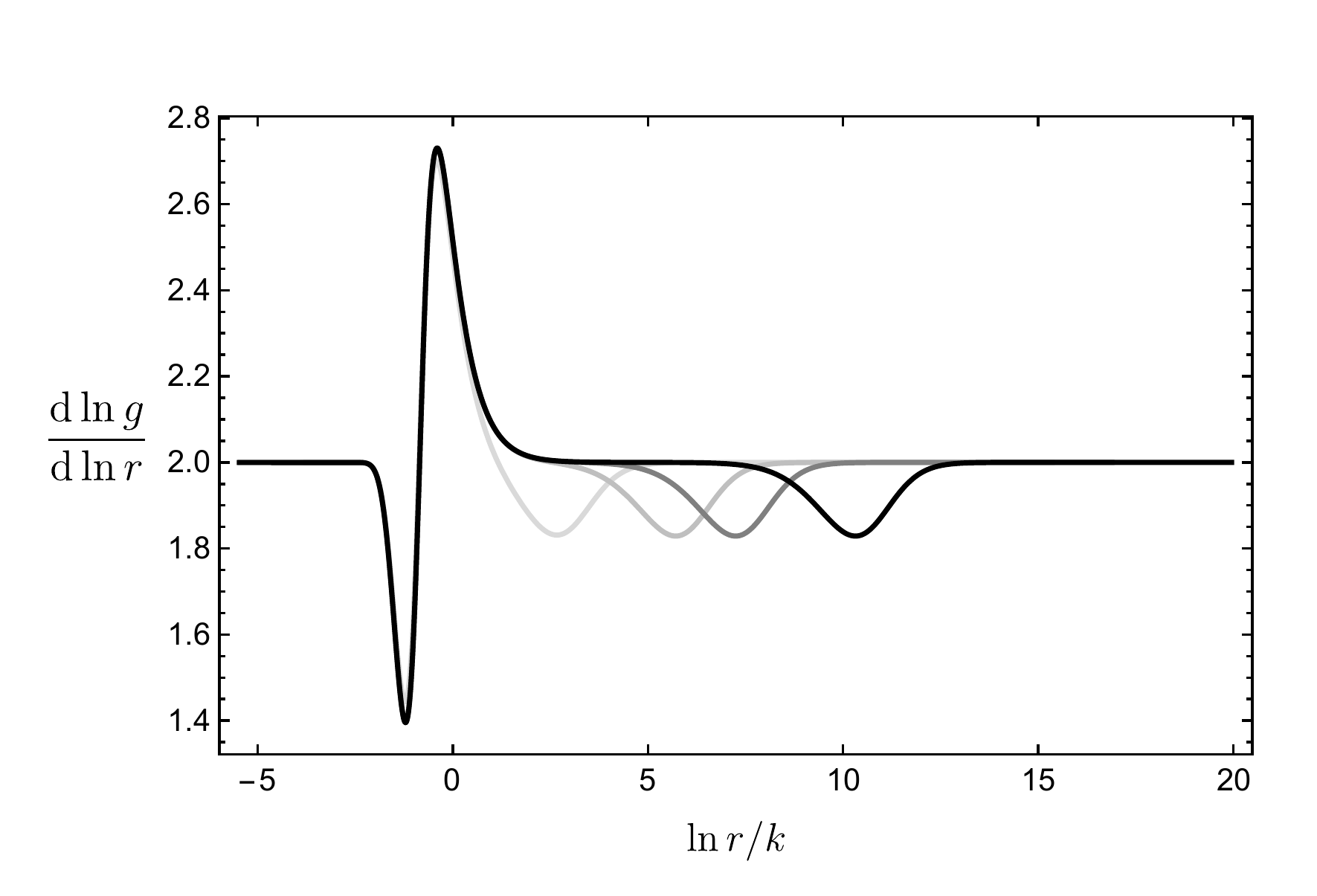}
\caption{Boomerang RG flows with $\Gamma/k^{3/2}= 10^2$ (lightest), $10^4, 10^5$ and $10^7$ (darkest) 
showing the build up of intermediate conformal invariance. The blue dashed line in the upper left plot shows the value of the scalar in the AdS$_5^c$ solution, with $\gamma_c =\sqrt{32/15}$. The plots clearly reveal an intermediate scaling region, dominated by the $AdS_5^c$ vacuum. The plots are for $m^2 =-15/4$ and $\xi = 675/512$.
  \label{fig:AdStoAdStoAds}}
\end{figure}

For sufficiently large $\Gamma/k^{3/2}$, the boomerang RG flow solutions start to exhibit 
intermediate scaling. Indeed, moving in from the UV, the solutions start to track the Poincar\'e invariant RG flow solutions
with $k=0$, for a range of the radial variable, including a region where the geometry approaches the $AdS_5^c$ solution,
before heading off back to the original $AdS_5^0$ solution in the deep IR. This behaviour is displayed for four representative
flows with
$\Gamma/k^{3/2}= 10^2, 10^4, 10^5$ and $10^7$
in figure \ref{fig:AdStoAdStoAds}. In this figure one can see the solution being dominated by
the $AdS_5^c$ solution for an intermediate range of $r/k$ which one can make parametrically large 
by increasing $\Gamma/k^{3/2}$.
Notice that for very large values of $\Gamma/k^{3/2}$, the function $\chi$ starts to take a kind of `sliding' form, in which the only significant
difference is the radial position in which $\chi$ increases to the first plateau. This behaviour gives rise to the asymptotic behaviour of $n$, but we have not been able to find a way to analytically extract the asymptotic value of $n$ as $\Gamma/k^{3/2}\to \infty$.

\section{Entanglement Entropy}\label{ententropy}
We now investigate how the intermediate scaling regime of the boomerang flows manifests itself in the entanglement entropy. 
Specifically we focus on calculating the holographic entanglement entropy for a ``strip geometry" of width $l$ in the $x$ direction \cite{Ryu:2006bv,Ryu:2006ef}.
We take a constant time slice and calculate the area, $\mathcal{A}$, of the minimal two-dimensional surface that is anchored
to the strip on the boundary. The entanglement entropy, $\mathcal{S}_\mathcal{A}$,
is then given by 
\begin{equation}
 \mathcal{S}_\mathcal{A} =4\pi \mathcal{A}\,.
\end{equation}
A standard calculation shows that the area can be expressed as
\begin{equation}
\mathcal{A} = \frac{2W^2}{r_\star\,^3}\int_{r_\star}^\Lambda\frac{r^5}{\sqrt{g\left(\frac{r^6}{r_\star\,^6}-1 \right)}}\mathrm{d}r\,,
\end{equation}
where $W^2$ is the area of the boundary strip in the $y,z,$ directions. We have integrated
from a minimum radial position of the surface at $r_\star$ to a UV cutoff $\Lambda$, which will eventually be taken to infinity.
We can relate $r_*$ to $l$ via the formula
\begin{equation}
l = 2\int_{r_\star}^\Lambda\frac{\mathrm{d}r}{r\sqrt{g\left(\frac{r^6}{r_\star\,^6}-1 \right)}}\,.
\end{equation}

At this point, setting $g=r^2/L^2$, we can easily recover the $AdS_5$ result of \cite{Ryu:2006bv,Ryu:2006ef}
\begin{equation}\label{eq:AdSMSA}
\frac{\mathcal{A}}{L^3} = \Lambda^2 \frac{W^2}{L^2} - b\left(\frac{W}{l}\right)^2\,.
\end{equation}
where $b=4\pi^{3/2}\left(\frac{\Gamma(\frac{2}{3})}{\Gamma(\frac{1}{6})}\right)^3$.
Note that in the limit $\Lambda\to\infty$ the area of the minimal surface displays the expected UV divergence. 
As this term is scheme dependent, it is natural to define the renormalised entanglement entropy, $\bar{\mathcal{S}}_\mathcal{A}$,
after subtracting off the UV divergence, as $ \bar{\mathcal{S}}_\mathcal{A} =4\pi (\mathcal{A}-\Lambda^2 W^2 L_0)$, and then take
$\Lambda\to \infty$.

The expression \eqref{eq:AdSMSA}
also motivates the definition of the so-called ``entropic $c$-function", defined for general bulk geometries via \cite{Nishioka:2009un}
\begin{equation}
C(l) = \frac{l^3}{W^2}\frac{\mathrm{d}\mathcal{S}_\mathcal{A}}{\mathrm{d}l}\,.
\end{equation}
In particular, for a bulk $AdS_5$ solution with radius $L$,  
we see that $C(l) =8\pi b L^3\approx 8.06 L^3$ and hence, as it is proportional to the $a$ central charge of the dual CFT,
provides a measure for the number of degrees of freedom in
the dual CFT. Furthermore, for Poincar\'e invariant RG flows, it has been shown that within two derivative gravity and with matter satisfying the null energy condition, $C(l)$ monotonically decreases along the RG flow \cite{Myers:2012ed}.

We now want to investigate $\bar{\mathcal{S}}_\mathcal{A}$ and $C(l)$ for the Poincar\'e invariant domain wall and the boomerang RG flows. A preliminary issue is to first check whether the flows introduce any additional UV divergences into the calculation of the minimal area. This would be the case if, for example, there was a constant term $g_0$ in the near boundary expansion $g\approx r^2 +g_0 +O(1/r)$. For the specific values of $m^2 =-15/4$, $\xi = 675/512$, it is straightforward to demonstrate that the Einstein equations require $g_0 = 0$. Thus, when constructing a renormalised entanglement entropy from the numerical data we need only to account for the quadratic ``area law" divergence of pure $AdS_5$. 

\begin{figure}
\centering
\includegraphics[scale=0.4]{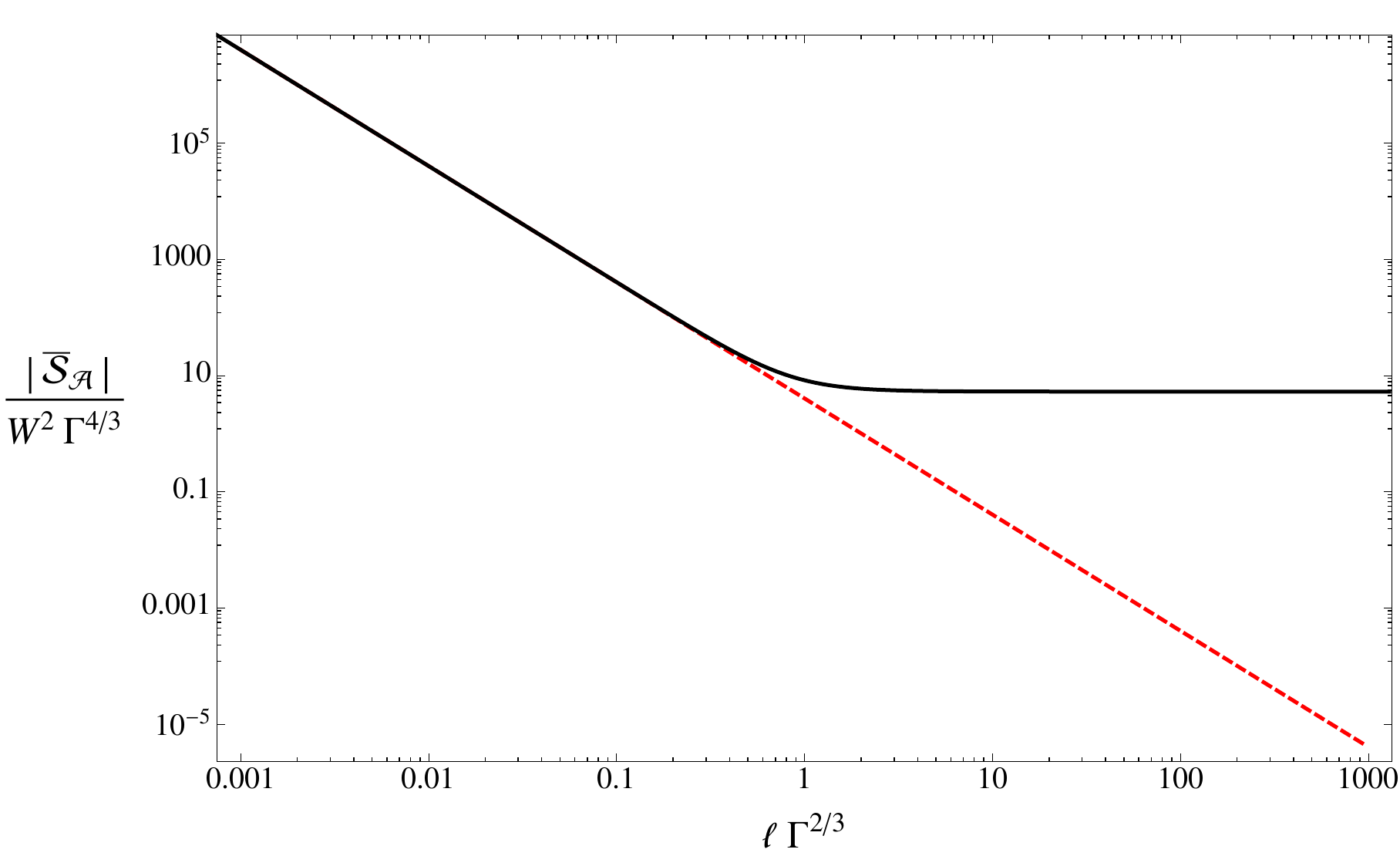}\quad
\includegraphics[scale=0.26]{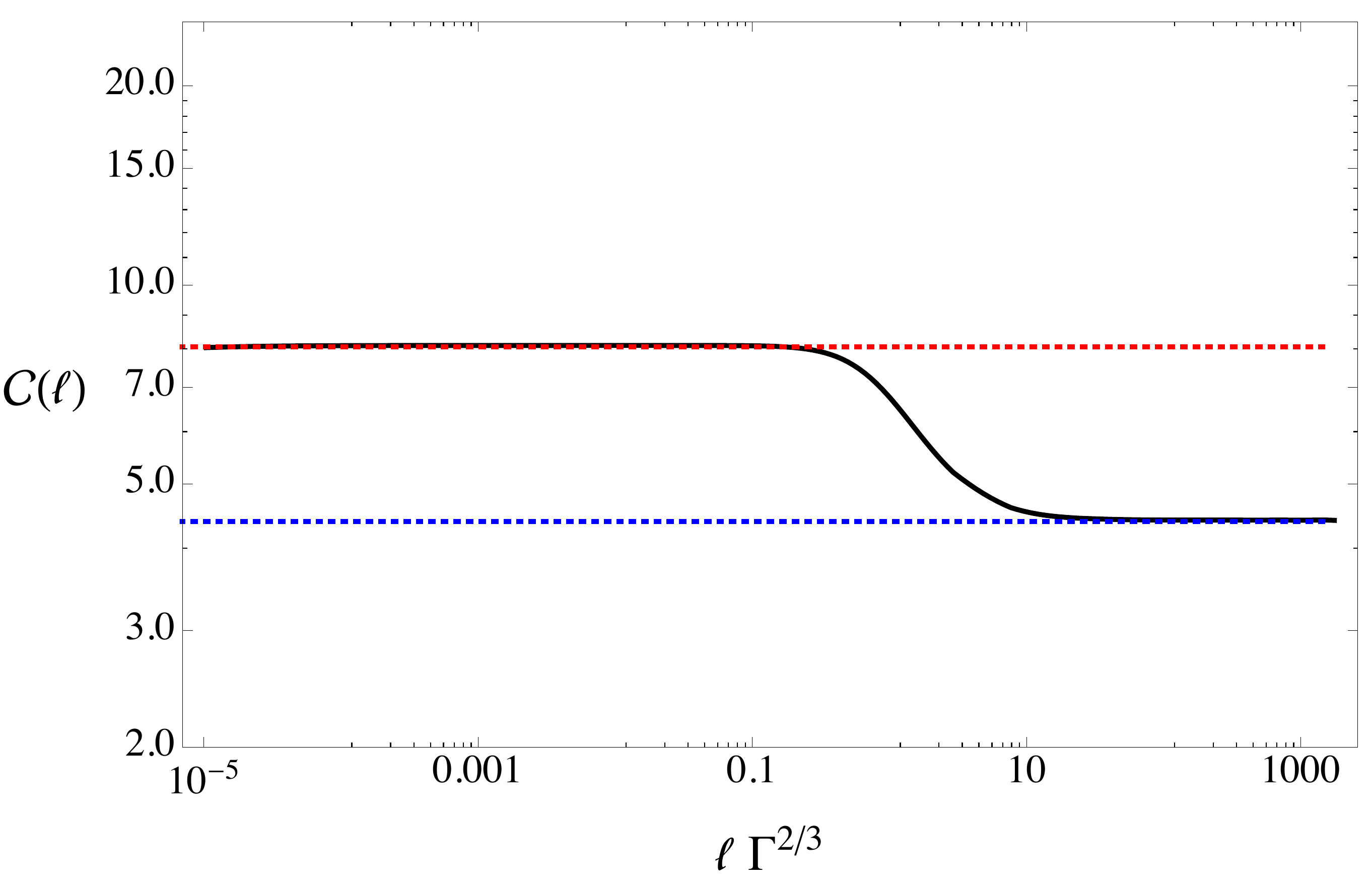}
\caption{The entanglement entropy $\bar{\mathcal{S}}_\mathcal{A}$ (left) and entropic $c$-function $C(l)$ (right) evaluated in the $AdS_5^0\to$ $AdS_5^c$ domain wall flow. The pure $AdS_5^0$ results are shown as dashed red lines, and agree excellently with the numerically computed quantities for small values of $l$.
The right plot shows that the entropic $c$-function ${C}(l)$ monotonically approaches the result for pure $AdS_5^c$ (lower dashed, blue line)
for large $l$. 
\label{fig:SandCAdS2AdS}}
\end{figure}

We next discuss the Poincar\'e invariant $AdS_5^0\to AdS_5^c$ domain wall solutions. For the usual values of $m^2 =-15/4$, $\xi = 675/512$, our results for
$\bar{\mathcal{S}}_\mathcal{A}$ and $C(l)$ are shown in figure \ref{fig:SandCAdS2AdS}. Notice that
for very large values of $l$, where the minimal surface is dipping deep into the $AdS_5^c$ geometry in the IR, $\bar{\mathcal{S}}_\mathcal{A}$ is not falling off
with increasing $l$ but instead asymptotes to a constant negative value. This behaviour was first observed for Poincar\'e invariant
domain wall solutions in \cite{Albash:2011nq}. Further insight into this phenomenon was also provided in \cite{Albash:2011nq} by showing how this result is expected at least in the case of very thin domain wall solutions. Figure \ref{fig:SandCAdS2AdS} also shows that the entropic $c$-function
$C(l)$ is a monotonically decreasing function of $l$, as expected,
interpolating between the $AdS_5^0$ result, $ C(l) \to 8\pi bL_0^3\sim 8.06$, as $l\to 0$ and the 
$AdS_5^c$ result, $ C(l) \to 8\pi bL_c^3\sim 4.39$, as $l\to \infty$.

\begin{figure}
\centering
\includegraphics[scale=0.41]{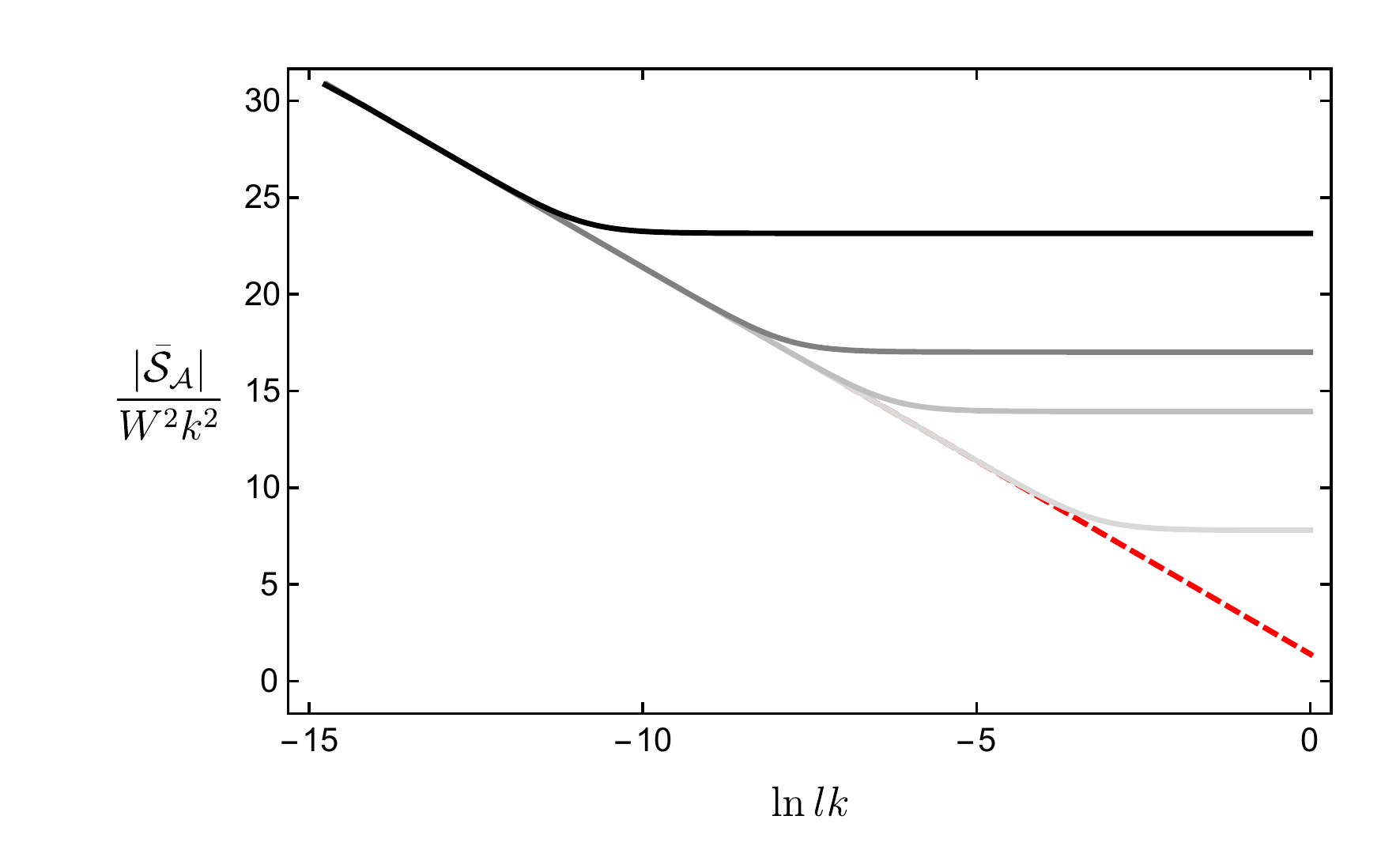}\quad
\includegraphics[scale=0.41]{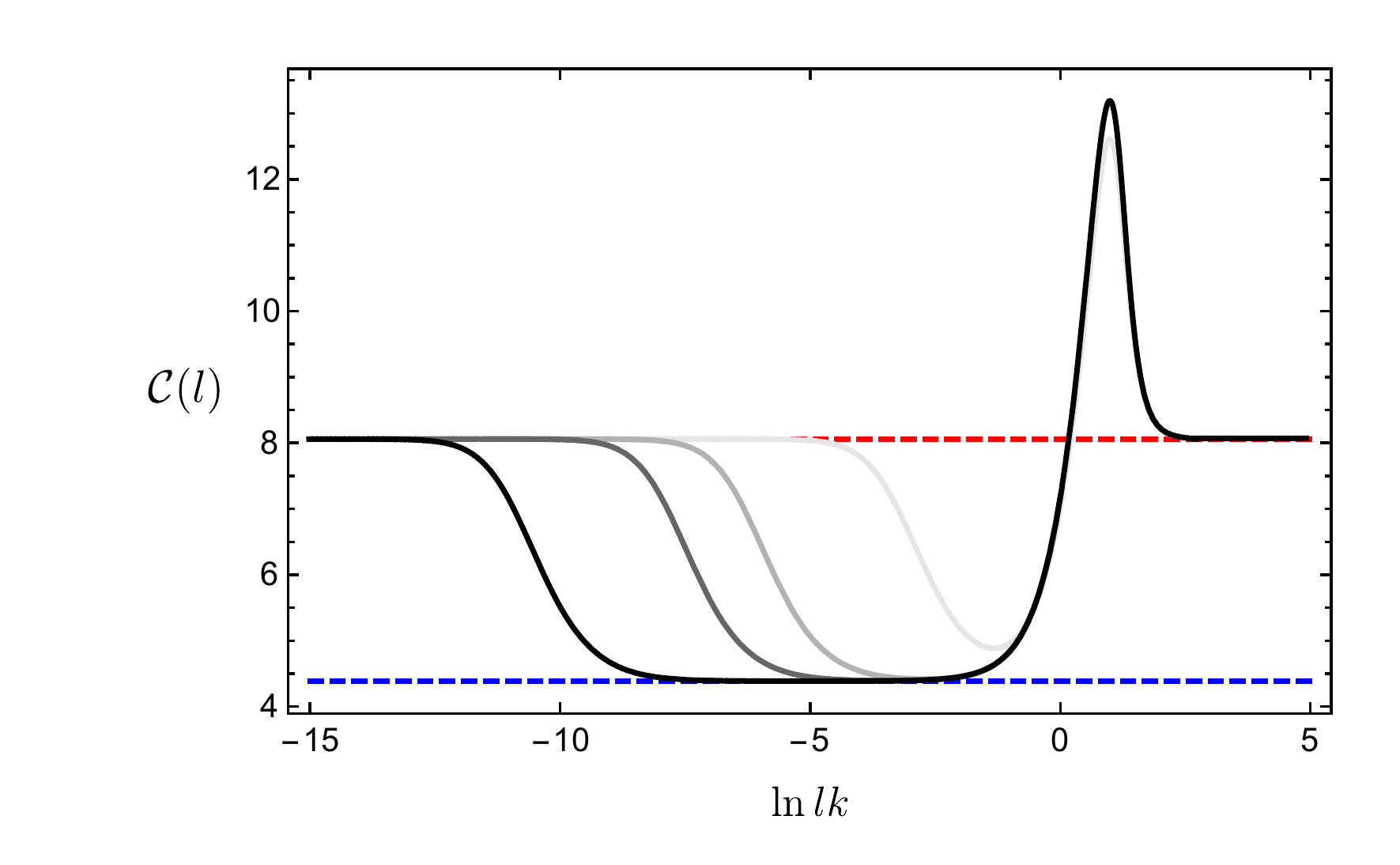}
\caption{The size of the entanglement entropy $|\bar{\mathcal{S}}_\mathcal{A}|$ (left) and the entropic $c$-function $C(l)$ (right)
evaluated for the boomerang RG flows for
$\Gamma/k^{3/2}= 10^2$ (lightest), $10^4, 10^5$ and $10^7$ (darkest), as in figure \ref{fig:AdStoAdStoAds}.
In the left plot, the pure AdS$_5^0$ result is a dashed red line.
In the right plot, the pure $AdS_5^0$ and $AdS_5^c$ results are given by the red and blue dashed lines, respectively.
\label{fig:SCAdS2AdS2AdS}}
\end{figure}

Finally, we turn to the boomerang RG flows, and our main results are shown in figure \ref{fig:SCAdS2AdS2AdS}.
For a given RG flow, labelled by $\Gamma/k^{3/2}$, the entanglement entropy looks qualitatively similar to the Poincar\'e invariant
domain wall solution: it is always negative and asymptotically approaches a negative constant for large $l$. 
For small values of  $\Gamma/k^{3/2}$ we can easily understand this behaviour using the perturbative solutions given in \eqref{eq:expg1},\eqref{eq:expg2}. Explicitly, starting from (\ref{eq:expg2}), writing $g = r^2 + \Gamma^2 \,u(r)+\ldots$ and expanding the integrand in the scalar amplitude $\Gamma$ yields
\begin{equation}\label{eq:AintEx}
\mathcal{A} = \mathcal{A}_{AdS} -\frac{W^2}{r_\star\,^3}\Gamma^2\int_{r_\star}^{\Lambda}\frac{r^2}{\sqrt{\frac{r^6}{r_\star\,^6}-1}}u(r)\,\mathrm{d} r\,,
\end{equation}
where $\mathcal{A}_{AdS}$ is the $AdS_5^0$ result, and the second term, which is finite in the limit $\Lambda\to\infty$, 
thus contains the entire finite contribution at large strip width.
To isolate the relevant part of the integral, we next turn our attention to the small $r_\star$ behaviour of the second term in (\ref{eq:AintEx}), since small $r_\star$ corresponds to large strip width in these backgrounds. The area in this limit is given by
\begin{align}
\mathcal{A} & =\, \mathcal{A}_{AdS} + \frac{W^2\,k^3}{4}\left(\frac{\Gamma}{k^{3/2}} \right)^2\int^\infty_{r_\star \to 0}\frac{e^{-2k/r}}{r^3}(2k-3r)\mathrm{d} r\,,
\end{align}
and hence the renormalised strip entanglement entropy evaluated in the perturbative RG flows asymptotes to 
\begin{equation}
\bar{\mathcal{S}}_\mathcal{A} \to -\pi W^2\,k^2\left(\frac{\Gamma}{k^{3/2}} \right)^2\,,
\end{equation}
at large strip width. This analytic prediction can be verified by comparing to the numerical results, and shows excellent agreement.
As we increase $\Gamma/k^{3/2}$, we find that $\bar{\mathcal{S}}_\mathcal{A}$ is always negative and from
figure \ref{fig:SCAdS2AdS2AdS} shows that it approaches an increasingly negative asymptotic value.

In figure \ref{fig:SCAdS2AdS2AdS} we see that in the boomerang RG flows the entropic $c$-function reveals additional interesting features. In the figure we have plotted $C(l)$ for the same value as in the boomerang flows of figure \ref{fig:AdStoAdStoAds}.
For small values of $l$ we see the expected $L_0^3$ behaviour of the $AdS_5^0$ solution with
$ C(l) \to 8\pi bL_0^3\sim 8.06$, as $l\to 0$.
At intermediate length scales, and for boomerang flows with intermediate scaling,
the function dips to a second plateau, much like in the Poincar\'e invariant domain wall, with
$ C(l) \to 8\pi bL_c^3\sim 4.39$ as one might have naively anticipated. Finally, far in the IR, the entropic $c$-function replicates the UV behaviour, as 
a consequence of the boomerang RG flow. Thus while $C(l)$
is certainly not a monotonic function along the RG flow, figure \ref{fig:SCAdS2AdS2AdS} shows that it does, nevertheless, 
provide a measure of the number of degrees of freedom in each of the three regions of the geometry where there is approximate conformal invariance, in the sense that for ranges of $l$ it approaches the result for the corresponding $AdS_5$ geometry.

\section{Boomerang flows with intermediate $AdS_2\times\mathbb{R}^3$ scaling and novel insulators}\label{qc}
Within the same class of models \eqref{eq:act}, but for a different range of the parameters $m^2$ and $\xi$, 
we now investigate another interesting framework in which instead of a second $AdS_5$ solution there is now 
an $AdS_2\times\mathbb{R}^3$ fixed point solution, which breaks translations. 
We will construct boomerang RG flows with locally quantum critical intermediate scaling, governed by the $AdS_2\times\mathbb{R}^3$
solution, as well as novel ground states that are thermal insulators. It is illuminating to construct the associated black hole solutions describing the systems at finite temperature and, for the specific values
$m^2 =-15/4$, $\xi = -1/4$, we find the phase diagram schematically shown in figure \ref{schempt}.

In this section (only) it will be convenient to use a slightly different radial variable than that of \eqref{eq:Qans}
and consider the ansatz 
\begin{align}\label{anstwo}
ds^2&=-Udt^2+U^{-1}d\rho^2+e^{2V}dx^\alpha dx^\alpha \,,\nn
z^{\alpha} &= \gamma e^{i k\, x^\alpha}\,,
\end{align}
with $U,V,\gamma$ functions of $\rho$. This ansatz can be used to construct both the RG flows and the black hole
solutions. We start by noting that for certain parameter ranges, we can construct $AdS_2\times\mathbb{R}^3$ solutions with $k\ne 0$, similar to the solutions discussed in the appendix of \cite{Donos:2013eha}. Specifically, we take
$U=\rho^2/L_{(2)}^2$, $V=0$, $\gamma=\gamma_{(2)}$ and
\begin{align}\label{ads2vals}
\gamma_{(2)}^2=\frac{\sqrt{12}}{\sqrt{-\xi}},\qquad k^2=\frac{\sqrt{-\xi}}{\sqrt{3}L_{(2)}^2},\qquad L_{(2)}^{-2}=8-\frac{m^2\sqrt{3}}{\sqrt{-\xi}}\,.
\end{align}
Clearly these solutions require models in which $\xi<0$. Now recall from \eqref{real} that the requirement that
there is a relevant scalar operator in the UV CFT with dimension $\Delta_0<4$, implies that $m^2<0$. From \eqref{irads} we see that
$m^2<0$ and $\xi<0$ are not compatible with having the second $AdS_5^c$ fixed point that we discussed in sections \ref{gensetup}-\ref{ententropy}.

We next consider the spectrum of deformations about the $AdS_2\times\mathbb{R}^3$ solution.
Considering perturbations of the form
\begin{align}\label{modeexp}
U=\frac{\rho^2}{L_{(2)}^2}(1+c_1\rho^\delta)\,,\qquad V=c_2 \rho^\delta\,, \qquad \gamma=\gamma_{(2)}(1+c_3 \rho^\delta)\,,
\end{align}
with $c_i$ constant,
then we find an unpaired mode with $\delta=-1$, with $c_2=c_3=0$, which simply corresponds to shifting 
$r$ by a constant in the solution. We also find a pair of modes with $\delta=-2,1$. The mode with  
$\delta=-2$ also has $c_2=c_3=0$ and is associated with heating up the solution. The mode with
$\delta=1$ has $c_1=(2/3+8L_{(2)}^2+\frac{\sqrt{3}}{\sqrt{-\xi}})c_3$, $c_2=-(8L_{(2)}^2+\frac{\sqrt{3}}{\sqrt{-\xi}})c_3$.
Finally there is
another pair of modes with $\delta=-\frac{1}{2}\pm\frac{1}{2}[1-\frac{64}{\sqrt{3}}L_{(2)}^2\sqrt{-\xi}]^{1/2}$ which have $c_2=0$ and 
$c_1$ related to $c_3$. Notice that there are BF violating modes in the $AdS_2$ solution when $\sqrt{-\xi}\ge \frac{\sqrt{3}}{16}(1+(1-4m^2)^{1/2})$.

In the remainder of this section\footnote{In appendix \ref{othervals} we will briefly consider models with $\xi=-675/512\sim -1.32$
which display some additional new features.} we will focus on models with parameters given by
\begin{align}\label{parvals}
m^2=-15/4\,,\qquad \xi=-1/4\,.
\end{align}
Choosing $m^2=-15/4$ (as in previous sections) implies that the UV CFT dual to the $AdS_5^0$ vacuum has a relevant scalar 
operator with dimension $\Delta_0=5/2$. In this case there are BF violating modes of the $AdS_2$ solution
when $\xi\le -75/256\sim-0.293$, and hence they are absent for \eqref{parvals}.
The numerical constructions of the solutions described in the following subsections are similar to those in previous sections and so we have
relegated some details to appendix \ref{appa}.

\subsection{RG flows}
We first consider RG flows which break spatial translations (i.e. with $k\ne0$),
starting from $AdS_5^0$ in the UV and going
to the $AdS_2\times\mathbb{R}^3$ solution in the IR, given by \eqref{ads2vals}. Note that with the values of the parameters given in \eqref{parvals} 
the value of the scalar field in the $AdS_2\times\mathbb{R}^3$ solution is $\gamma_{(2)}=48^{1/4}\sim 2.63$. From the point of view of
the IR, we use the $\delta=1$ mode mentioned below \eqref{modeexp} to shoot out from the $AdS_2\times\mathbb{R}^3$ solution, as described in appendix \ref{appa}. For future reference we note that this mode
is associated with an irrelevant operator in the CFT dual to $AdS_2\times\mathbb{R}^3$ with $\Delta=2$.
These RG flows exist for a specific value of the dimensionless deformation parameter which is numerically found to be at $\Gamma/k^{3/2}=\bar\Gamma$, with 
\begin{align}
\bar\Gamma\sim 19.37\,.
\end{align}

\begin{figure}
\centering
\includegraphics[scale=0.41]{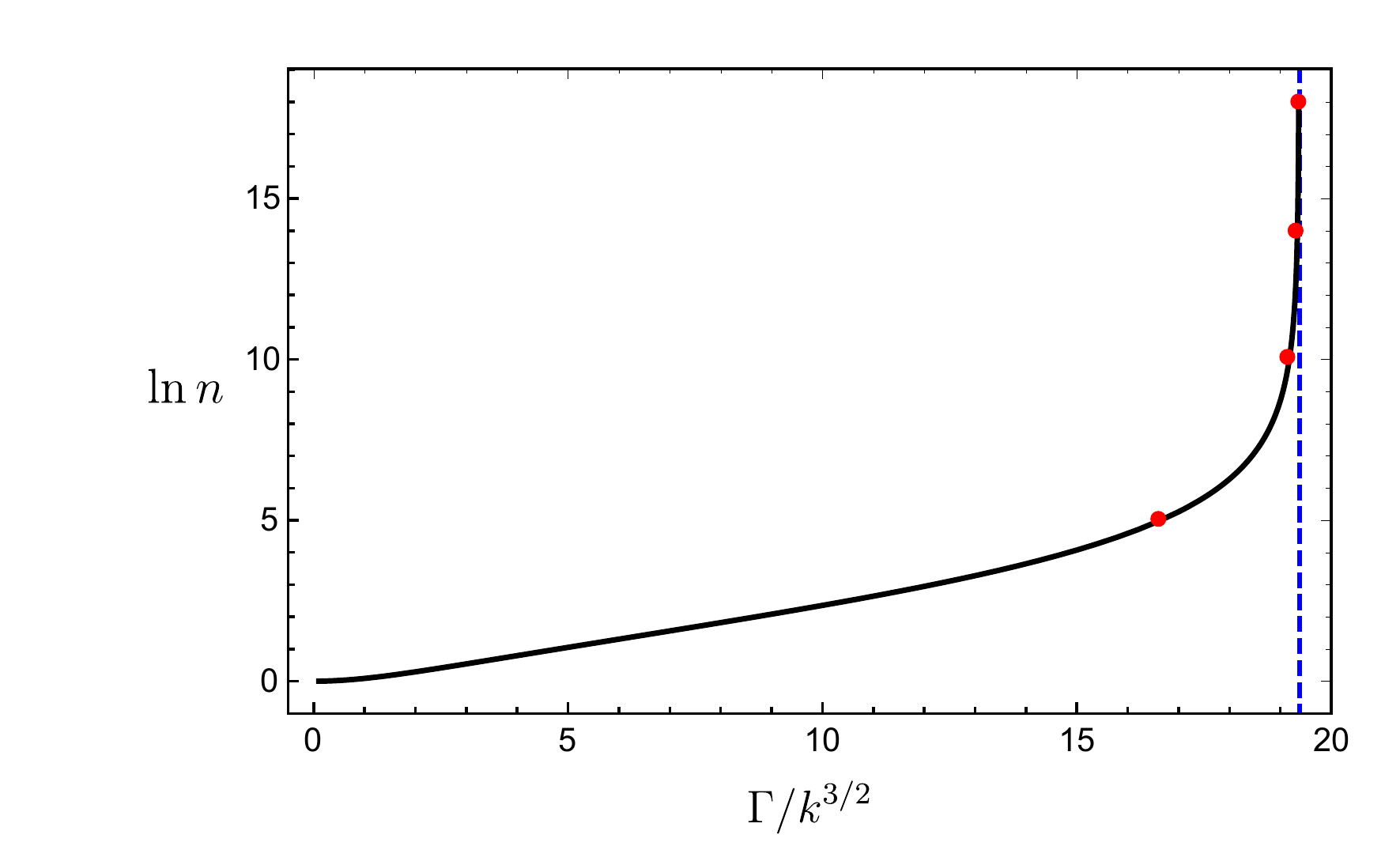}
\includegraphics[scale=0.41]{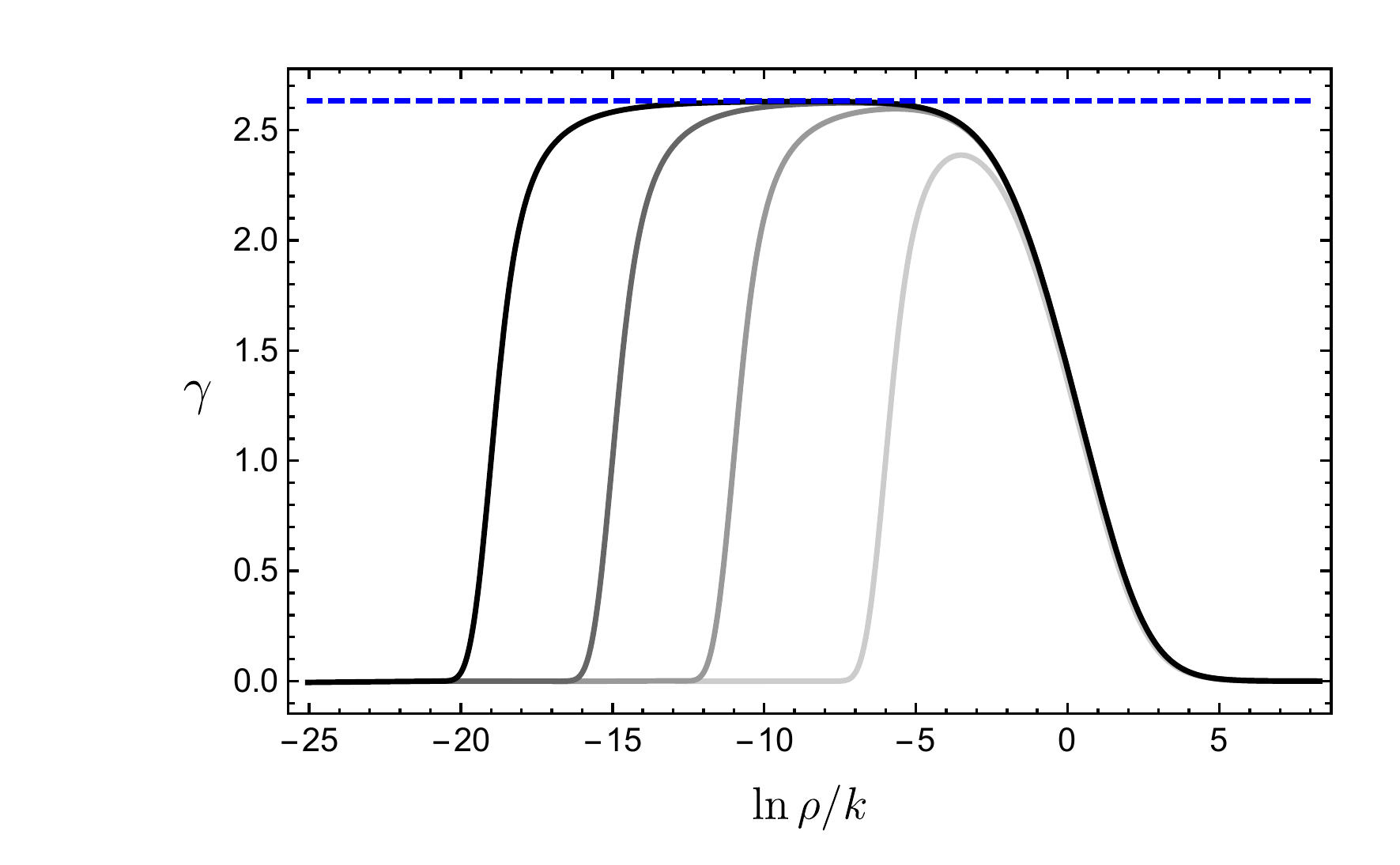}
\caption{Boomerang RG flows for $m^2 =-15/4$ and $\xi =-1/4$. The left plot shows the refractive index $n$, defined
by \eqref{iref}, as a function of the deformation parameter
$\Gamma/k^{3/2}$, for the one parameter family of boomerang RG flows that exist in the finite range 
$0\le \Gamma/k^{3/2}\le \bar\Gamma$. For the special value $\Gamma/k^{3/2}=\bar\Gamma\sim 19.37$ (dashed vertical line) we have the $AdS_5^0$ to $AdS_2\times\mathbb{R}^3$ RG flow. As $\Gamma/k^{3/2}$ approaches $\bar\Gamma$
there is a build up of an intermediate scaling regime, determined by the $AdS_2\times\mathbb{R}^3$ solution, as displayed
by the radial behaviour of the scalar function for $\Gamma/k^{3/2} =16.65$, $19.17$, $19.33$ and $19.36$ 
(red dots on the left plot and light grey to dark grey on the right plot) and we have also marked
the value of the scalar in the $AdS_2\times\mathbb{R}^3$, $\gamma_{(2)}=48^{1/4}$, by a horizontal dashed line.
  \label{figboom1}}
\end{figure}

We next consider the boomerang RG flows starting from $AdS^0_5$ in the UV and ending up at the same $AdS^0_5$ in the IR. In the coordinates we are using the index of refraction is now given by
\begin{align}\label{iref}
n=e^{V_{IR}-V_{UV}}\,.
\end{align}
In figure \ref{figboom1} we have plotted some features of the one parameter family of boomerang RG flows that we have constructed numerically, which, interestingly, exist in the finite range $0\le \Gamma/k^{3/2}\le \bar\Gamma$.
We find that as $\Gamma/k^{3/2}$ approaches $\bar\Gamma$ the boomerang RG flows start to build up an intermediate
scaling regime governed by the $AdS_2\times\mathbb{R}^3$ solution. 
We have also calculated the holographic free energy for these RG flows and we find that 
as $\Gamma/k^{3/2}\to\bar\Gamma$ we have $ T^{tt}/k^4\to 1614$, 
which is in excellent numerical agreement with the
value of the free energy that we directly obtain for the $AdS_5^0$ to $AdS_2\times\mathbb{R}^3$ RG flow.

While satisfying, this analysis does not reveal what happens for RG flows with $\Gamma/k^{3/2}> \bar\Gamma$. It turns out
that these flows are singular in the far IR. In order to elucidate what is going on, we use the standard technique of
constructing finite temperature black holes and then cooling them down to low temperatures. As we will see this will
also reveal interesting features at finite $T$ for $\Gamma/k^{3/2}< \bar\Gamma$.

\subsection{Black hole solutions and thermal insulators}
\begin{figure}[h!]
\centering
\includegraphics[scale=0.4]{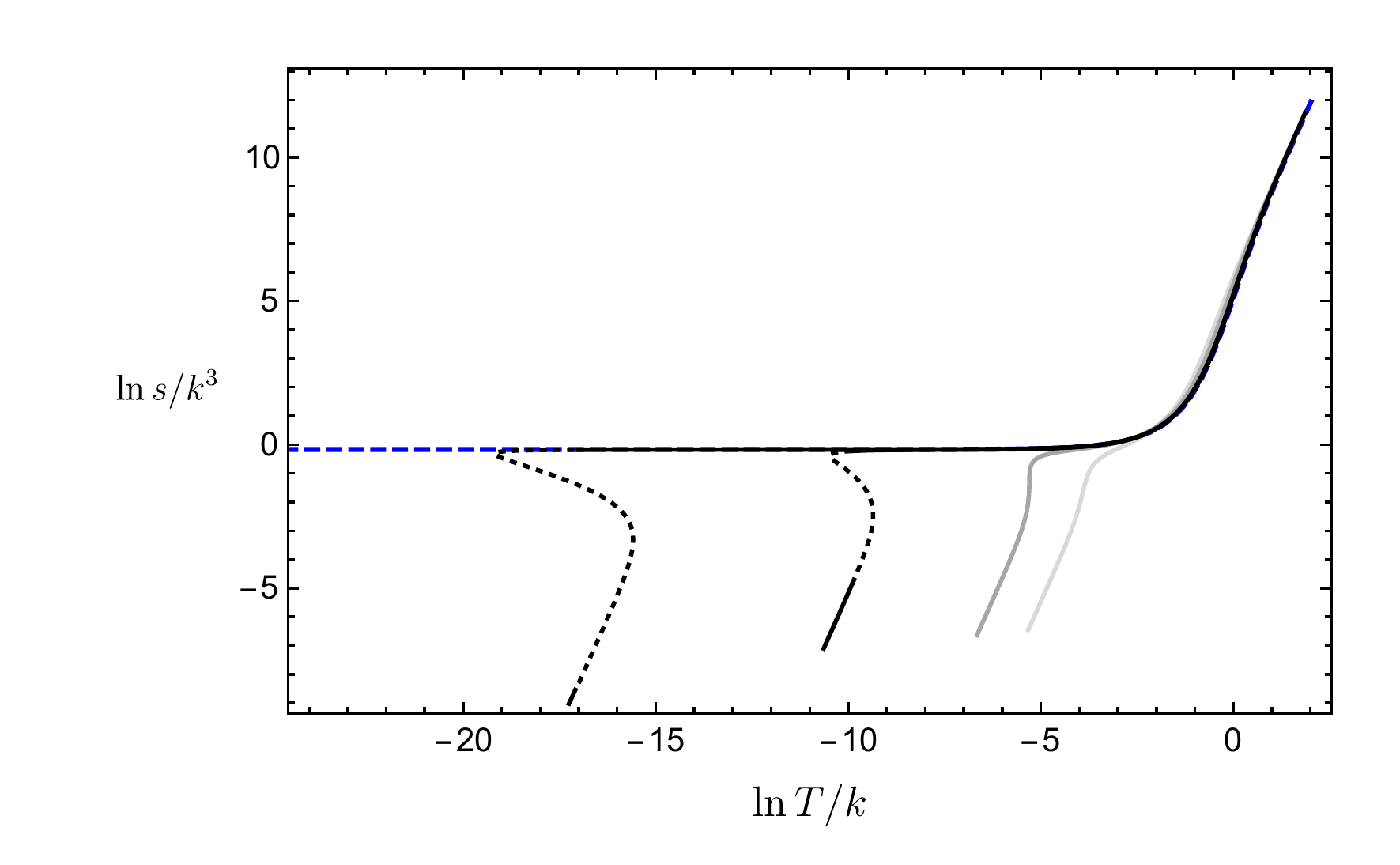}\quad
\includegraphics[scale=0.4]{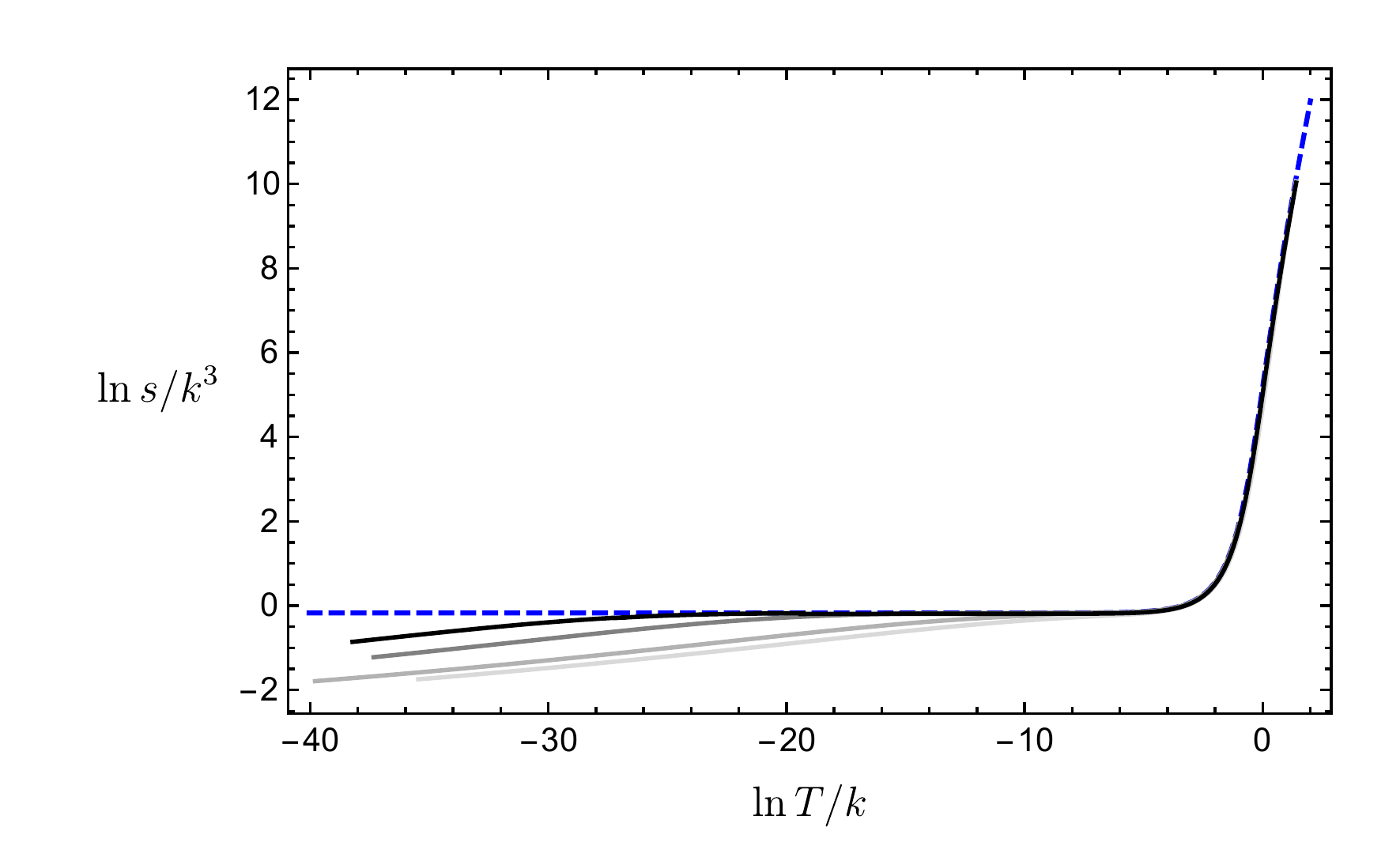}
\caption{Plots of the entropy density $s$ as a function of $T$ for black hole solutions with
various deformation parameters and for models with $m^2 =-15/4$ and $\xi =-1/4$. 
The left plot is for $\Gamma/k^{3/2}\le\bar\Gamma\sim 19.37$, namely $5.5$, $10.3$, $18$, and $19.3$ (light grey to dark grey). 
The dashed blue line in both plots is for $\Gamma/k^{3/2} =\bar\Gamma$.
For $\Gamma/k^{3/2} =19.33$ and $19.36$ we have used dotted lines on the curves to indicate the first order phase transition.
As $T\to 0$, for $\Gamma/k^{3/2}<\bar\Gamma$ the black holes approach the boomerang RG flows, with $s\sim T^3$, 
and when $\Gamma/k^{3/2}=\bar\Gamma$ they approach the $AdS_5^0$ to $AdS_2\times\mathbb{R}^3$ RG flow with, from \eqref{ads2vals},
$s \sim 0.84 k^3$. 
The right plot is for $\Gamma/k^{3/2}\ge \bar\Gamma$, namely, $19.41$, $19.75$, $25$ and $35$ (dark grey to light grey). As $T\to 0$
the black holes have $s\to 0$, but not as a power of $T$. At $T\to 0$ they approach insulating ground states. In both plots
the intermediate scaling regimes can be seen.
  \label{entngs}}
\end{figure}

It is straightforward to numerically construct black hole solutions for arbitrary values of $\Gamma/k^{3/2}$ using the IR expansion
as given in appendix \ref{appa}. We first consider the black hole solutions in the range $0\le \Gamma/k^{3/2}< \bar\Gamma$, where
the boomerang RG flows exist. 
In the subrange $0\le \Gamma/k^{3/2}\le \Gamma_C$, below a critical value $\Gamma_C$, we find that the black hole solutions can be cooled down, uneventfully, and they smoothly approach the boomerang RG flows at $T=0$. In particular, we find that as $T\to 0$ the entropy, $s$, goes to zero with
$s\sim T^3$.
However, in the range $\Gamma_C<\Gamma/k^{3/2}<\bar\Gamma$ there is a first order phase
transition at finite temperature. The immediate signal for this behaviour can be seen in the entropy versus
temperature plots becoming multivalued, as displayed in figure \ref{entngs}. Furthermore, by
calculating the free energy for the black holes in this range as a function of $T$, which 
display the characteristic swallowtail behaviour, we can determine the thermodynamically preferred black holes,
again shown in \ref{entngs}.

When $\Gamma/k^{3/2}=\Gamma_C$
the line of first order phase transitions ends in a second order critical point with
$\Gamma_C\sim 10.5$.
When $\Gamma/k^{3/2}=\bar\Gamma$
the line of first order phase transitions smoothly ends at $T=0$ at the $AdS_5^0$ to $AdS_2\times\mathbb{R}^3$
RG flow solution.
This behaviour is summarised in figure \ref{schempt}.

We next construct black hole solutions with $\Gamma/k^{3/2}> \bar\Gamma$. In this case we find no evidence of any phase transitions at finite $T$. 
When $\Gamma/k^{3/2}$ is close to $\bar\Gamma$, as the temperature is lowered the solutions build up a finite temperature behaviour that is governed by the $AdS_2\times\mathbb{R}^3$ solution, before heading off to the new $T=0$  ground states. This is displayed in 
the behaviour of entropy versus temperature plots shown in figure \ref{entngs} and also
in figure \ref{schempt}. The plots in figure \ref{entngs} also reveal some important features
of the new $T=0$ ground states. Firstly, as $T\to 0$ we have $s\to 0$. However, unlike other $s=0$ ground states which break translations \cite{Donos:2012js,Donos:2014uba,Gouteraux:2014hca,Donos:2014oha},
we find that the entropy density is not vanishing as a power law with $T$. Indeed we find that
$d(\ln s)/d(\ln T)$ is decreasing as $T\to 0$. The $T=0$ ground states are certainly singular: for example the value of the scalar field 
at the horizon diverges as $T\to 0$.

\begin{figure}[h!]
\centering
\includegraphics[scale=0.41]{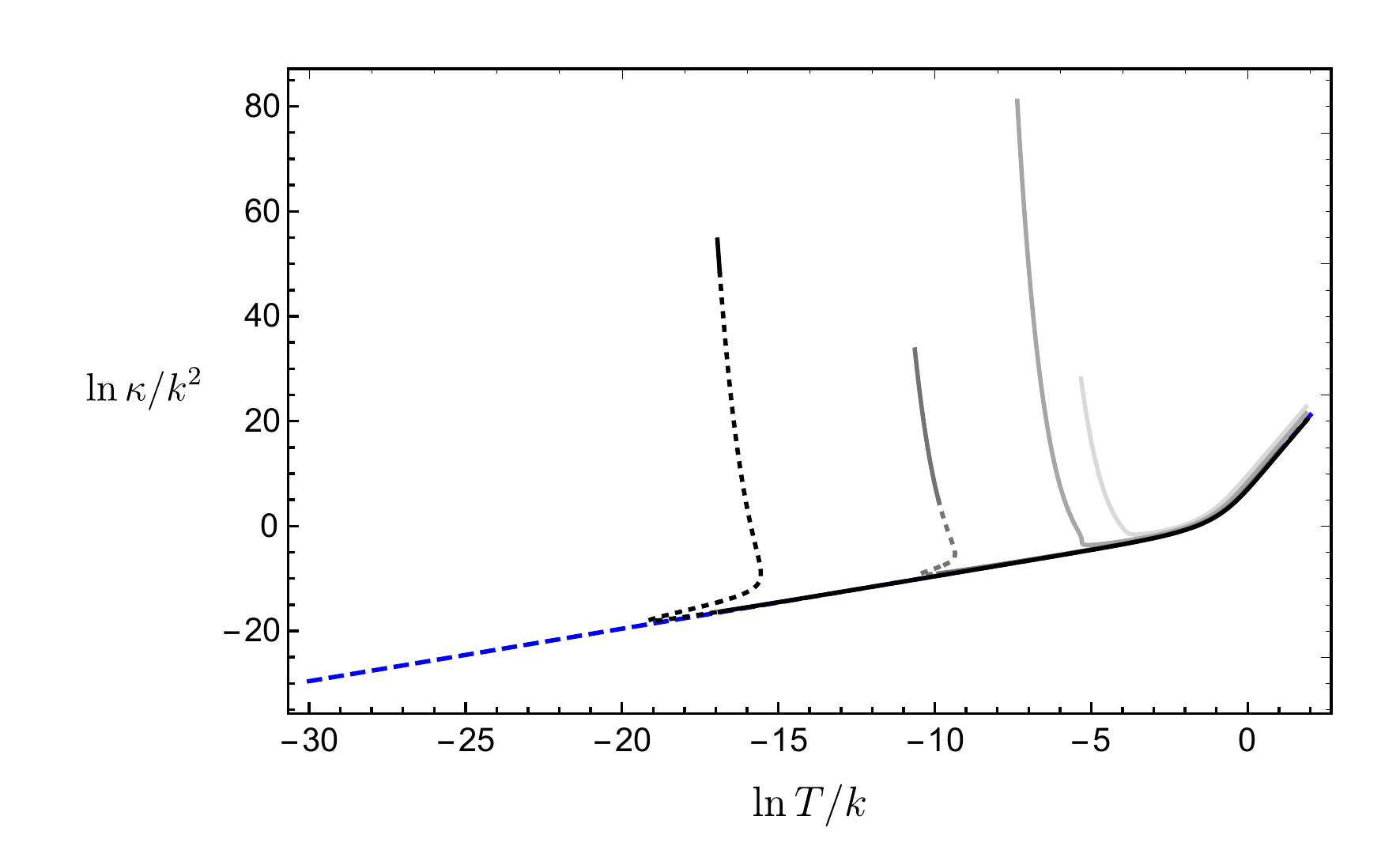}\quad
\includegraphics[scale=0.41]{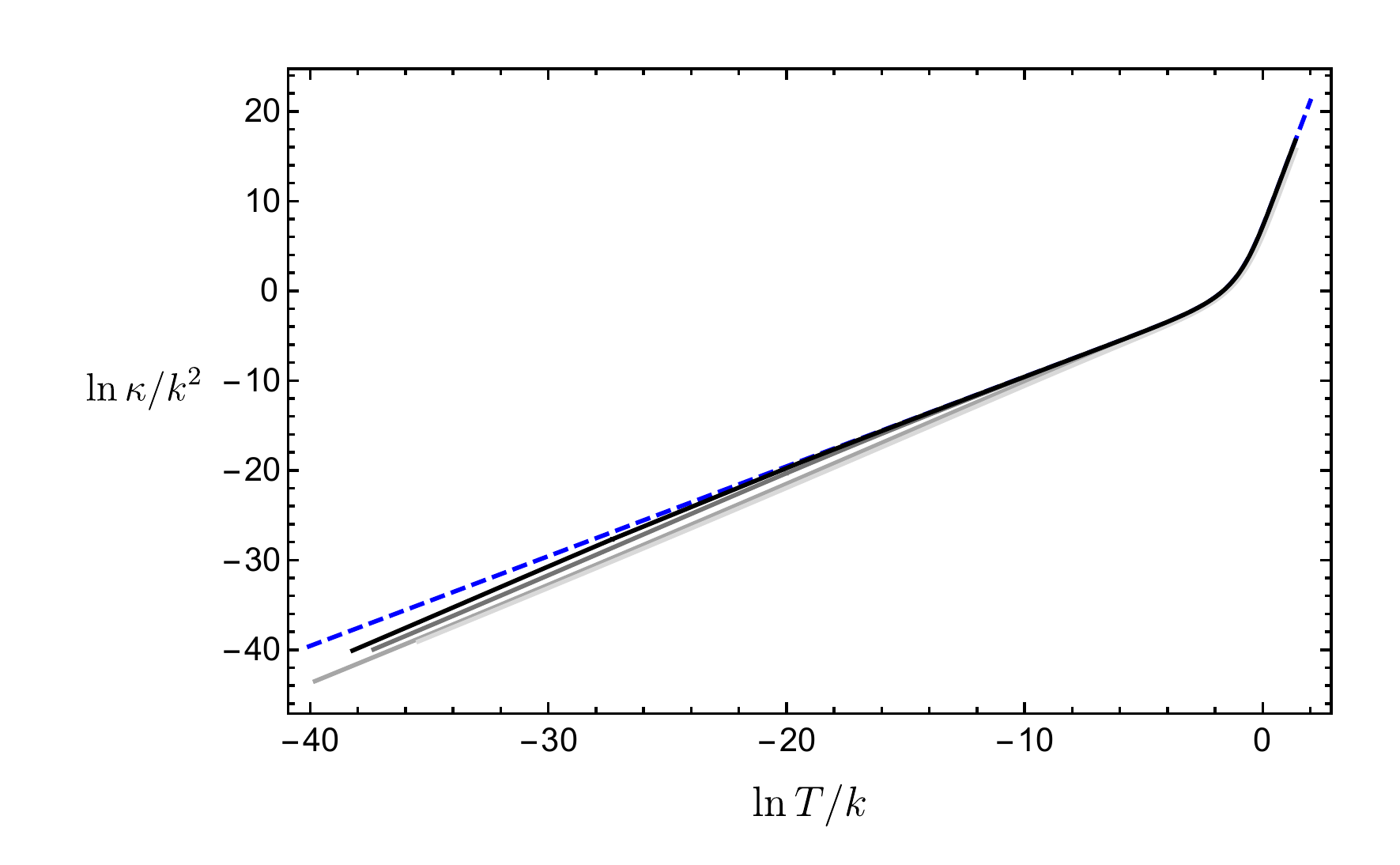}
\caption{Plots of the temperature dependence of the thermal conductivity $\kappa$ for the black hole solutions
constructed for $m^2 =-15/4$ and $\xi =-1/4$. The left plot is for $\Gamma/k^{3/2}\le\bar\Gamma\sim 19.37$, namely 
$5.5$, $10.3$, $18$, and $19.3$
(light grey to dark grey), as in figure \ref{entngs}, and we see thermal conducting behaviour with $\kappa\to\infty$  
as $T\to 0$. The dashed blue line in both plots is for $\Gamma/k^{3/2} =\bar\Gamma$ with $\kappa\sim T$ as $T\to 0$.
The right plot is for $\Gamma/k^{3/2}\ge \bar\Gamma$, namely, $19.41$, $19.75$, $25$ and $35$ (dark grey to light grey), as in figure \ref{entngs},
and we see thermal insulating behaviour with $\kappa\to 0$ as $T\to 0$.  
  \label{kappadiff}}
\end{figure}

To determine some additional properties of the $T=0$ ground states for $\Gamma/k^{3/2}> \bar\Gamma$ we can calculate the
thermal DC conductivity matrix, $\kappa^{ij}$. For general holographic lattices this can be calculated by solving a Stokes flow at the horizon \cite{Donos:2015gia}. In fact,
for this case we can use the results presented in \cite{Donos:2014cya} which showed that for all of the black hole solutions that we have constructed
we have $\kappa^{ij}=\kappa \delta^{ij}$ with 
\begin{align}\label{kapathor}
\kappa=\frac{4\pi s T}{\gamma_H^2 k^2}\,,
\end{align}
where $\gamma_H$ is the value of the scalar field at the black hole horizon.
Plotting this as a function of $T$ we find the behaviour shown in the right panel of figure \ref{kappadiff}, clearly revealing that, for
$\Gamma/k^{3/2}> \bar\Gamma$, $\kappa\to 0$ as $T\to 0$. By examining the $d \ln \kappa/d\ln T$ we deduce that $\kappa$ is
not going to zero as a power law, in line with the radial behaviour of the metric mentioned previously. 
This thermal insulating behaviour arises because $sT/\gamma_H^2\to 0$ as $T\to 0$ i.e. the number of degrees of freedom available to transport heat, captured by $s$, is going to zero rapidly enough as $T\to 0$.

For $\Gamma/k^{3/2}= \bar \Gamma$, associated with the
$AdS_5^0$ to $AdS_2\times\mathbb{R}^3$ RG flow, we have $\kappa\sim T$.
For $\Gamma/k^{3/2}< \bar\Gamma$ the $T=0$ ground states are the boomerang RG flows which have translationally invariant
horizons and hence $\kappa(T)\to\infty$ as $T\to 0$, 
as we see in figure \ref{kappadiff}. We can be slightly more precise about this behaviour, generalising arguments
in \cite{Donos:2014gya,Donos:2016zpf}, essentially by heating up the boomerang RG flow. By considering the $AdS$-Schwarzschild black hole we deduce that the location of the black hole horizon is related to the temperature via $\rho_+=\pi T$. Next, by considering \eqref{expsdwap} we deduce that we can 
obtain the renormalisation of length scales\footnote{For the coordinates we are using, $\bar L$ is the same as the index of refraction for the boomerang RG flow given in \eqref{iref}.} in the boomerang RG flow, $\bar L$, by taking the following limit of the black hole solutions:
$\bar L=\lim_{T/k\to 0}[ e^V|_{\rho=\rho_+}/(\pi T)]$.  Using \eqref{expsdwap} and \eqref{kapathor}, we then  anticipate that as $T/k\to 0$, the thermal conductivity blows up as an exponential multiplied by a factor of $T^7$. We have verified that this is the case for several  branches of black holes with  $\Gamma/k^{3/2}< \bar\Gamma$.

Finally, although not displayed in figure \ref{kappadiff}, as $T\to\infty$ we find that $\kappa\to T^7$. 
This behaviour can be understood using a similar argument, as given\footnote{Note that the $T\to\infty$ expression for $\kappa$ given in eq. (3.30)
of \cite{Donos:2014cya} is only valid for non-vanishing charge density.}
in section 3.3 of \cite{Donos:2014cya}, to show that the high temperature behaviour of
$\kappa$ is given by $\kappa\to T^{2(6-\Delta)}$, where $\Delta=5/2$ is the scaling deformation of the operator dual to the complex
scalar fields in \eqref{eq:lag} for the UV $AdS_5^0$ vacuum. 

\subsection{Diffusion and butterfly velocity}
The black hole solutions that we have constructed all explicitly break translation invariance in the dual field theory. On general
grounds \cite{Donos:2017gej,Donos:2017ihe}, the black holes necessarily have a quasinormal hydrodynamic mode associated with diffusion of 
heat. From the results of 
\cite{Donos:2017ihe} this mode has 
a diffusion constant, $D$, which governs the dispersion relation of the mode, which can be obtained via the Einstein relation $D=\kappa/c$, where $c\equiv T\partial s/\partial T$ is the specific
heat (holding the deformation parameter $\Gamma/k^{3/2}$ fixed). Since we have already calculated, numerically, both $\kappa(T)$ and $s(T)$ 
it is therefore straightforward to obtain $D(T)$.

We next consider the calculation of the butterfly velocity $v_B$. This can be obtained by studying the construction of a shockwave
geometry on the black hole horizon \cite{Shenker:2013pqa,Maldacena:2015waa}. For the class of metrics we are considering, from \cite{Blake:2016wvh} we have
\begin{align}
v^2_B=\frac{4\pi T}{6[e^{2V}\dot V]_H}\,.
\end{align}
We now consider the possibility that we have a relationship of the form
\begin{align}\label{dvbee}
D=E\frac{v^2_B}{2\pi T}\,,
\end{align}
where we are interested in the low temperature behaviour of the dimensionless quantity $E(T)$

\begin{figure}[h!]
\centering
\includegraphics[scale=0.4]{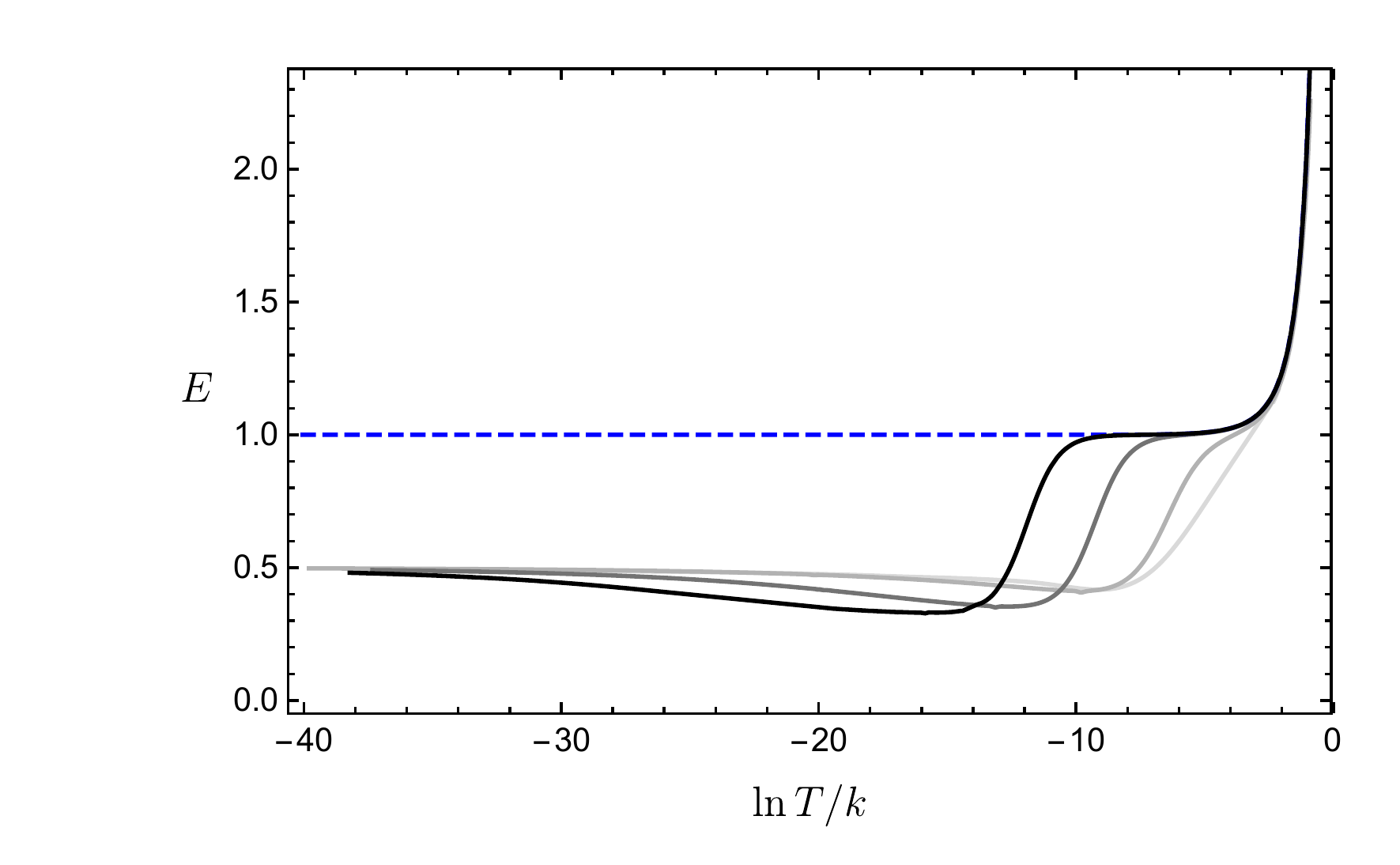}
\caption{Behaviour of the ratio of the the thermal diffusion constant to the butterfly velocity, $E\equiv D/(v^2_B/2\pi T)$,
as a function of $T$ for the black hole solutions constructed for $m^2 =-15/4$ and $\xi =-1/4$ and $\Gamma/k^{3/2}\ge \bar\Gamma$. 
For $\Gamma/k^{3/2}=\bar\Gamma$, dashed blue line, we see $E\to 1$ as $T\to 0$. For the thermal insulators with 
$\Gamma/k^{3/2}$ equal to $19.75$, $25$ and $35$ (dark grey to light grey), as in the right panels of
figure \ref{entngs} and \ref{kappadiff}, we see that $E\to 1/2$ as $T\to 0$.
  \label{ET}}
\end{figure}

For $\Gamma/k^{3/2}=\bar\Gamma$ we see from figure \ref{ET} that as $T\to 0$ we
have $E\to 1$, 
in agreement with the results of \cite{Blake:2016jnn}, where we recall that the $AdS_5^0$ to $AdS_2\times\mathbb{R}^3$ domain wall solution is
being driven by an irrelevant operator in the locally quantum critical CFT dual to $AdS_2\times\mathbb{R}^3$ with scaling dimension $\Delta=2$. For $\Gamma/k^{3/2}< \bar\Gamma$ (not shown in figure \ref{kappadiff})
we have the boomerang RG 
flows and the $T=0$ ground states are $AdS_5^0$. Since these ground states are 
translationally invariant,
the diffusion mode is absent: as $T\to 0$ we have $\kappa\to \infty $, $D\to \infty$ and $E \to \infty$.
However for boomerang RG flows with
intermediate scaling governed by the $AdS_2\times\mathbb{R}^3$ solution, we find that there
is a range of temperatures where $E\sim 1$.

Finally, of most interest, we consider the low temperature behaviour of $E$ for
for $\Gamma/k^{3/2}> \bar\Gamma$, associated with the thermally insulating ground states.
Remarkably, to good numerical accuracy, we find that $E\to 0.5$ as $T\to 0$, as shown in figure \ref{kappadiff}.

\section{Final comments}
Within a Q-lattice framework, in sections \ref{gensetup}-\ref{ententropy}
we constructed simple holographic solutions which describe boomerang RG
flows from a CFT in the UV, deformed by spatially dependent 
relevant operators, to the same CFT in the IR.
For large enough deformation the solutions approach an intermediate scaling regime with a new conformal symmetry
appearing at intermediate scales, which is governed by another $AdS_5$ solution.

The main features of this construction, combined with insights obtained in \cite{Donos:2017ljs},
indicate that within a bottom up framework there is significant freedom to construct `designer boomerang flows'.
In particular, suppose that we want to construct a boomerang flow from a holographic fixed point in the UV to the same fixed
point in the IR. This fixed point does not have to be conformal and could be, for example, of Lifshitz type. Suppose
also that we want an intermediate scaling regime governed by some other holographic geometry which could be $AdS$, 
Lifshitz or even hyperscaling violation form, which doesn't break translations. Then one looks for a gravitational model in which there is a standard
RG flow from the UV fixed point to the intermediate scaling geometry driven by a relevant deformation, which we suppose is driven by bulk scalar fields. In addition we demand that the gravitational model allows for a Q-lattice ansatz in which the bulk scalar fields
depend on the spatial coordinates. Then, much as in this paper, there should be RG flows parametrised by a dimensionless parameter
of the form $\Gamma/k^\alpha$, where $k$ is the characteristic wave number of the spatial deformation and $\Gamma$ characterises the strength of the deformation of the relevant operator. For small $\Gamma/k^\alpha$ one expects
a boomerang RG flow as the perturbative mode rapidly dies out in the IR\footnote{An exception would be if the perturbative expansion generates zero mode terms at higher orders which change the IR. Such behaviour can be eliminated by imposing discrete symmetries.}. If the boomerang RG flow exists for large values of $\Gamma/k^\alpha$ then the desired intermediate scaling will also appear. In simple models the latter feature should be present, but on the other hand it is not guaranteed. For example, if there were additional fixed point solutions in the model, then one may be driven away from the boomerang flows by a quantum phase transition at some value of $\Gamma/k^\alpha$ before intermediate scaling is seen.

In section \ref{qc} we also constructed boomerang flows for models in which there was an $AdS_2\times\mathbb{R}^3$ solution which breaks translation invariance. In this case, for a specific value of $\Gamma/k^{3/2}=\bar\Gamma$ there is an RG flow solution from $AdS_5^0$ in the UV to $AdS_2\times\mathbb{R}^3$ in the IR. 
We found that as $\Gamma/k^{3/2}$ approached $\bar\Gamma$ from below, the boomerang RG flow solutions build up an increasingly large $AdS_2\times\mathbb{R}^3$ intermediate scaling region. {\it A priori}, it is unclear what might happen for $\Gamma/k^{3/2}>\bar\Gamma$.
However, by constructing finite temperature black holes, in addition to finding an interesting line of first order phase transitions
for $\Gamma/k^{3/2}<\bar\Gamma$, we found a new class of thermally insulating ground states for $\Gamma/k^{3/2}>\bar\Gamma$.
A particularly interesting feature of these new ground states is that $s(T) \to 0$ but not as a power law. We also showed, numerically, 
that these ground states
exhibit a simple relationship between the thermal diffusion constant and the butterfly velocity of the form $D=E v^2_B/(2\pi T)$ with 
$E(T)\to 0.5$
as $T\to 0$. It would certainly be interesting to have a better analytic understanding of these ground state solutions, which should also allow
us to confirm this result for $E(T)$. 
An interrelated point would be to obtain a better understanding of these insulating ground states in the limit of large $\Gamma/k^{3/2}$.
A more general point is that it is not at all clear how one can directly construct such novel ground state solutions, without power law behaviour, in holography.
What we have shown in this paper is that for at least one class of models, analysing models with boomerang RG flows can reveal them.

In the above discussion we have been considering deformations associated with relevant operators of the UV fixed point.
However, we note that boomerang RG flows with intermediate scaling and within a Q-lattice framework do not require the deformations to be associated with relevant operators: indeed the examples in \cite{Donos:2016zpf}
were driven by marginal operators. It is also worth emphasising that boomerang RG flows do not require Q-lattices and can also arise for what are called inhomogeneous lattices. For example, 
in \cite{Chesler:2013qla} boomerang RG flows were driven by a deformation involving a spatially varying chemical potential of CFT in $d=3$ of the form $V\cos kx$.
For these deformations with $k=0$ and $V\ne 0$, we have the standard zero temperature 
AdS-Reissner-Nordstrom black hole, which approaches $AdS_2\times \mathbb{R}^2$ in the IR. 
It is therefore natural to conjecture that for sufficiently large $V/k$ the boomerang RG flows could have an intermediate scaling
regime approaching $AdS_2\times \mathbb{R}^2$. It would be of interest to examine this in more detail, and more generally,
analyse boomerang RG flows for other inhomogeneous lattices.

\section*{Acknowledgements}
We thank Clifford Johnson for a helpful discussion.
The work of JPG and CR is supported by the European Research Council under the European Union's Seventh Framework Programme (FP7/2007-2013), ERC Grant agreement ADG 339140. The work of JPG is also supported by STFC grant ST/P000762/1, EPSRC grant EP/K034456/1, as a KIAS Scholar and as a Visiting Fellow at the Perimeter Institute.

\appendix

\section{Some details for section \ref{qc}}\label{appa}
For the ansatz used in \eqref{anstwo} the equations of motion are given by
\begin{align}
0 &= \dot U + U\left(2 \dot V -\frac{\dot \gamma^2}{2 \dot V}\right)-\frac{1}{\dot V}\left(4-\frac{1}{2}(e^{-2V}k^2+m^2)\gamma^2-\frac{\xi}{3}\gamma^4\right)\,,\nn
0 &= \ddot\gamma +\dot \gamma\left(\frac{\dot U + 3 U\dot V}{U}\right)-\gamma\left(\frac{e^{-2V} k^2 +m^2}{U}\right)-\frac{4\xi}{3 U}\gamma^3\,,\nn
0 &= \ddot V +\dot V^2 +\frac{1}{2}\dot \gamma^2\,.
\end{align}

Restricting to $m^2=-15/4$ we have the following UV expansion as $\rho\to\infty$, 
\begin{align}\label{uvcons2}
U &= (\rho+\rho_+)^2(1-\frac{3\Gamma^2}{8(\rho+\rho_+)^3}+\frac{M}{(\rho+\rho_+)^4}+\dots )\,,\nn
e^{2V}&=(\rho+\rho_+)^2(1- \frac{3 \Gamma^2}{8 (\rho+\rho_+)^3}-\frac{5\Gamma\Gamma_2}{8 (\rho+\rho_+)^4}+ \dots  )\,,\nn
\gamma &= \frac{\Gamma}{(\rho+\rho_+)^{3/2}}+\frac{\Gamma_2}{(\rho+\rho_+)^{5/2}}+\frac{\Gamma k^2}{2(\rho+\rho_+)^{7/2}}+\dots\,.
\end{align}
where the parameter $\rho_+$, obtained by shifting the radial coordinate is included to conveniently place the IR in the RG flows
at $\rho=0$.

For the boomerang RG flows the IR expansion, at $\rho\to 0$, is given by
\begin{align}\label{expsdwap}
U &= \rho^2\left[1+ \frac{3}{16}e^{-\frac{2k}{\rho c_V}}\left(2 c_V \frac{k^2}{\rho^2}+ c_V^2\frac{k}{\rho}\right)\left(\frac{c_\gamma}{k^{3/2}}\right)^2+...\right]\,,\nn
e^{2V} &= \rho^2c_V^2\left(1-\frac{1}{16}e^{-\frac{2k}{\rho c_V}}\left(4\frac{k^3}{\rho^3}+3c_V^2\frac{k}{\rho}+3 c_V^3\right)\left(\frac{c_\gamma}{k^{3/2}}\right)^2+..\right)\,,\nn
\gamma &= \frac{k^{3/2}}{\rho^{3/2}}e^{-k/\rho c_V}\left(\frac{c_\gamma}{k^{3/2}}\right)+\dots\,,
\end{align}
fixed by two parameters $c_\gamma$ and $c_V$.
For the RG flow from $AdS_5^0$ to $AdS_2\times \mathbb{R}^3$ solution we use the IR expansion 
\begin{align}\label{irdf}
U&=\frac{\rho^2}{L_{(2)}^2}(1+c_1\, \rho+\ldots)\,,\nn
e^{2V}&=c_V(1+2c_2 \,\rho+\ldots)\,,\nn
\gamma&=\gamma_{(2)}(1+c_3\,\rho+\ldots)\,,
\end{align}
with the $c_i$ as given for the $\delta=1$ mode in section \ref{qc}, and hence, with $c_V$, \eqref{irdf} depends on two free constants.

Finally, for the black hole solutions we use the expansion as $\rho\to 0$ given by
\begin{align}\label{horexp}
U &= 4\pi T \rho - \rho^2\frac{1}{48} \left(96+45\gamma_H^2-8\xi\gamma_H^4 -36\gamma_H^2(k/V_H)^2\right)+...\,,\nn
e^{2V} &=V_H^2+\rho \frac{V_H^2}{48\pi T}\left(96+45\gamma_H^2-8\xi\gamma_H^4-12\gamma_H^2(k/V_H)^2\right) +...\,,\nn
\gamma &=\gamma_H+ \rho\frac{\gamma_H}{48\pi T}\left( -45+16\xi \gamma_H^2+   12(k/V_H)^2\right)+...\,.
\end{align}

In order to calculate the holographic energy of the domain wall solutions, we need to calculate the holographic stress tensor.
To do this, we need to supplement the bulk action \eqref{eq:act}
with boundary terms including the usual Gibbons-Hawking term and a counter-term action given by
$S_{ct}=\frac{1}{16\pi G}\int d^4 x\sqrt{-\gamma}(6+\sum_\alpha \frac{3}{4}z^\alpha\bar z^{\alpha}+\dots)$, where $\gamma_{ij}$ is the 
pull back of the bulk metric to the regulating UV boundary
and the neglected terms are not important for the calculation of interest. After a little calculation we find that the energy density of the dual field
theory is given in terms of the UV expansion of \eqref{uvcons2} as
\begin{align}
  T^{tt}=-3(M+\Gamma\Gamma_2)\,.
\end{align}
In order to determine the thermodynamically preferred black holes we also need to calculate the free energy
density, $w$, and we have 
\begin{align}
w=  T^{tt}-Ts\,.
\end{align}

\subsection{RG flows and Black holes for $m^2=-15/4$, $\xi=-675/512$}\label{othervals}
In section \ref{qc} we focussed on models with $m^2=-15/4$ and $\xi=-1/4$. Here we briefly
discuss models with $m^2=-15/4$, $\xi=-675/512\sim -1.32$ which exhibit some interesting different behaviour.
It is worth noting that for these values, the $AdS_2\times\mathbb{R}^3$ solution given in \eqref{ads2vals}
has a BF violating mode (see the discussion below \eqref{modeexp}) and this seems to be at least partially responsible for the differing behaviour.

\begin{figure}
\centering
\includegraphics[scale=0.4]{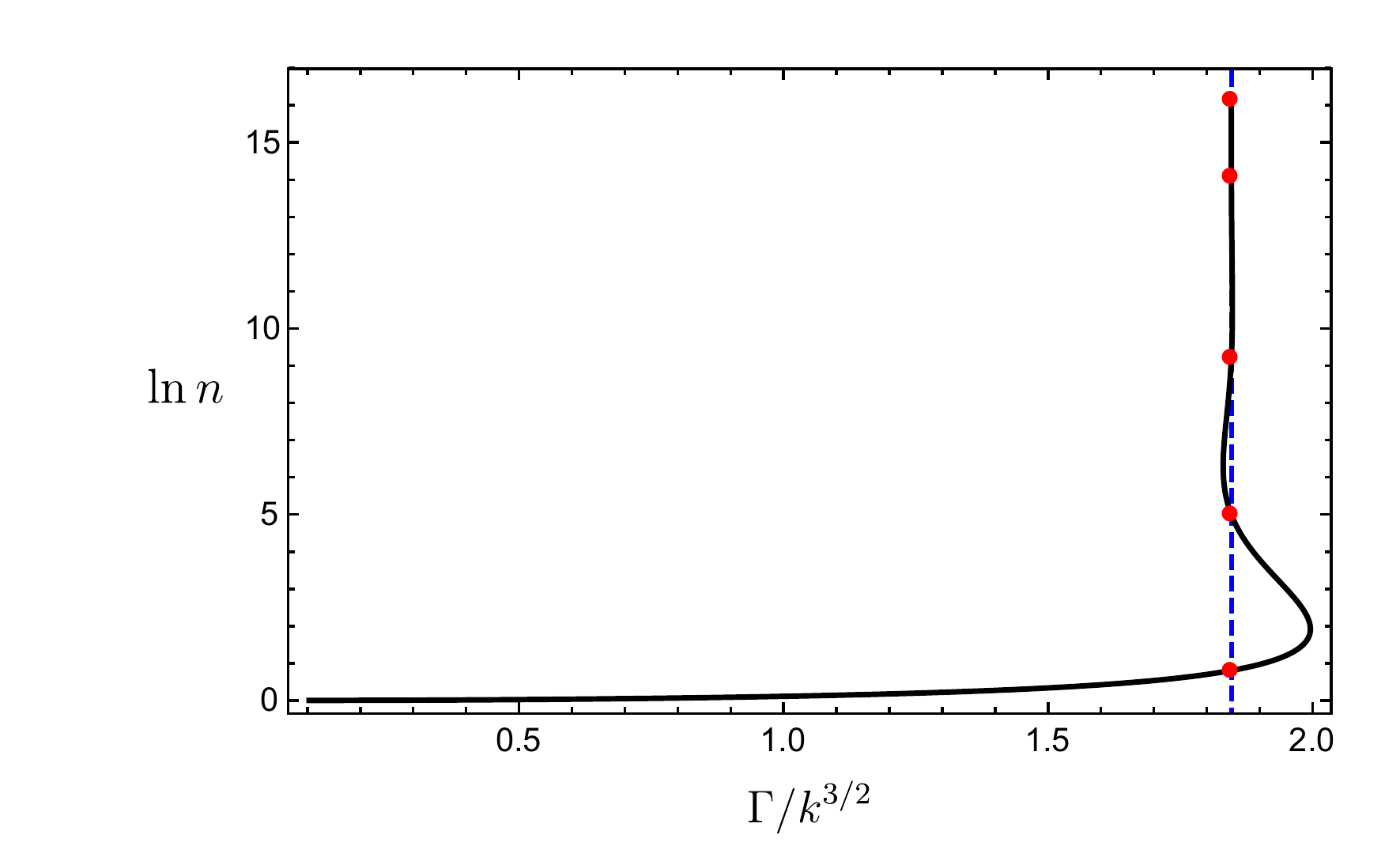}\quad
\includegraphics[scale=0.4]{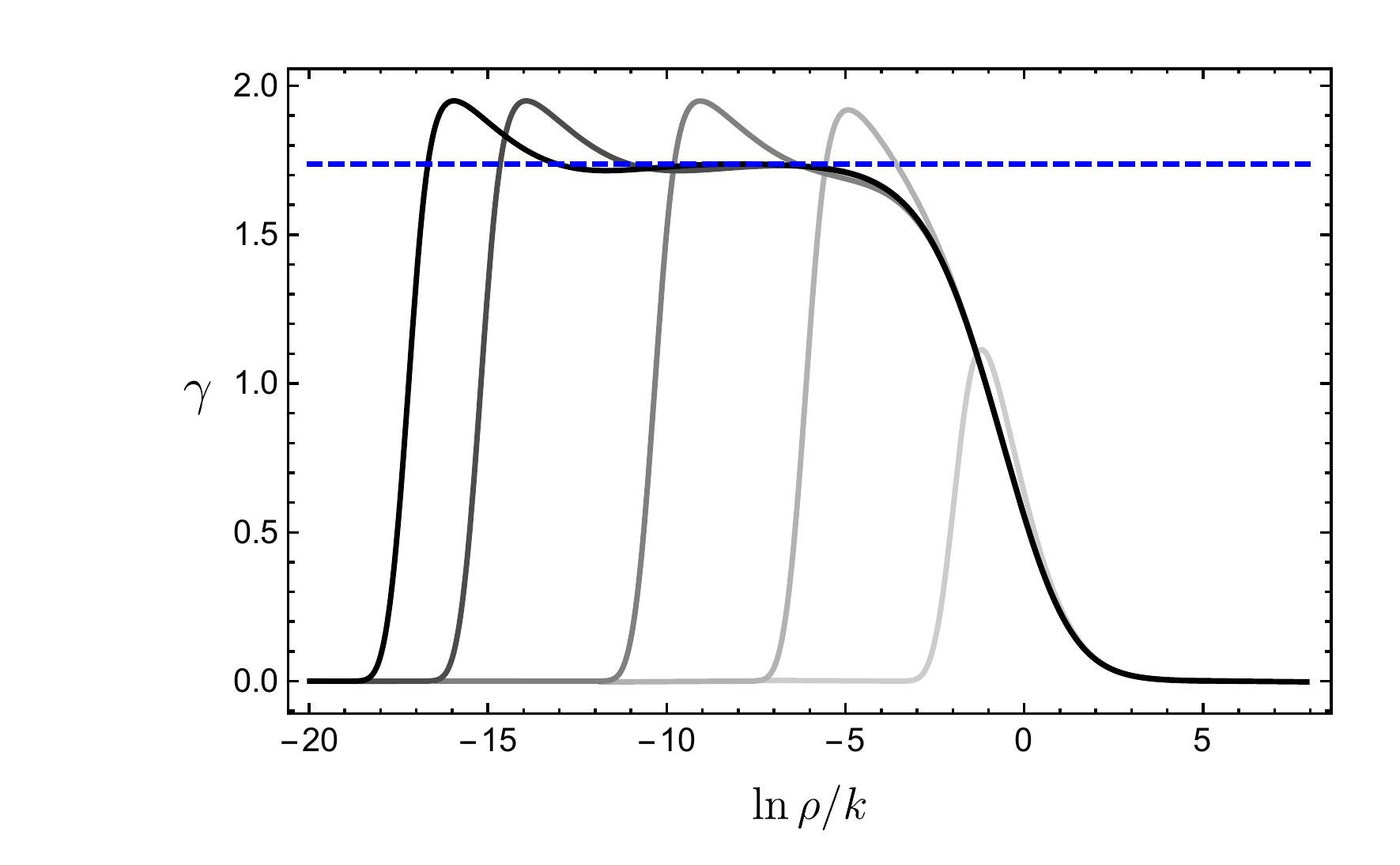}\\
\includegraphics[scale=0.4]{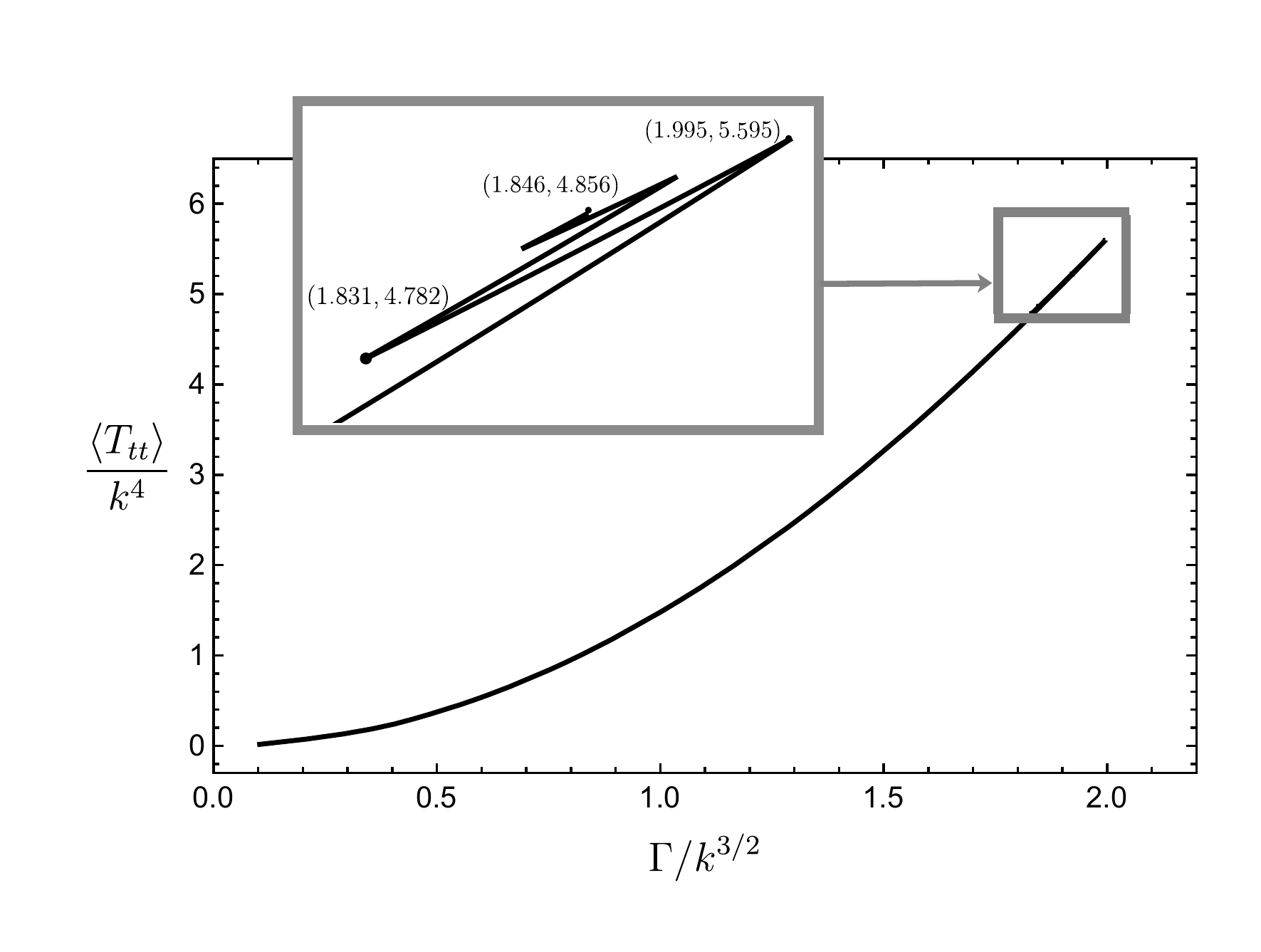}
\caption{Boomerang RG flows for $m^2 =-15/4$ and $\xi =-675/512$. The upper left plot shows the refractive index $n$, defined
by \eqref{iref}, as a function of the deformation parameter
$\Gamma/k^{3/2}$, for a one parameter family of boomerang RG flows. For the special value $\Gamma/k^{3/2}=\bar\Gamma\sim 1.8466$ (dashed vertical line) there is also an $AdS_5^0$ to $AdS_2\times\mathbb{R}^3$ RG flow. Near $\Gamma/k^{3/2}= \bar\Gamma$ the boomerang flows are not
uniquely specified by the value of $\Gamma/k^{3/2}$. As $n$ becomes large the boomerang flows build up a large intermediate scaling regime dominated by
the $AdS_2\times\mathbb{R}^3$ solution: for several boomerang flows with $\Gamma/k^{3/2}\sim 1.8466$, 
denoted by red dots, we have plotted
the radial behaviour of the scalar function in the upper right plot, with light grey to dark grey associated with increasing $n$, and the dashed horizontal line indicates the constant value of $\gamma$ in the 
$AdS_2\times\mathbb{R}^3$ fixed solution. The bottom plot shows the value of the energy for the boomerang flows and  we see that
for values of $\Gamma/k^{3/2}$ where there is non-uniqueness, it is the solution with the smallest value of $n$ that is preferred (the lightest grey in the upper right plot), 
and hence the intermediate $AdS_2\times\mathbb{R}^3$ scaling is frustrated.
  \label{othercase}}
\end{figure}

The $AdS_5^0$ to $AdS_2\times\mathbb{R}^3$ RG flow now exists for the special value $\Gamma/k^{3/2}= \bar\Gamma\sim 1.8466$. We can also construct boomerang RG flows from $AdS_5^0$ to $AdS_5^0$ for a finite range of $\Gamma/k^{3/2}$ but unlike when
$m^2=-15/4$, $\xi=-1/4$, and surprisingly, this range now extends further than $\bar\Gamma$ as shown in figure \ref{othercase}.
In particular, there is now a range of $\Gamma/k^{3/2}$ where the boomerang RG flows are not uniquely determined by
the UV deformation parameter $\Gamma/k^{3/2}$, but instead by the refractive index $n$.

These are perhaps the first examples of holographic RG flows which have the same fixed point solution in the IR
and yet they are not uniquely specified by their UV deformation data\footnote{Examples of holographic RG flows which are 
not uniquely specified by their UV data but with {\it different} IR fixed point solutions arise in many situations including 
in the context of the 
$T\to 0$ limit of spontaneously broken phases.}. Furthermore, for the specific value $\Gamma/k^{3/2}=\bar\Gamma$
there is also the $AdS_5^0$ to $AdS_2\times\mathbb{R}^3$ RG flow solution, giving rise to an additional non-uniqueness
for this specific value of the UV deformation. 
As $n\to\infty$ we find that $\Gamma/k^{3/2}\to \bar\Gamma$. Furthermore, as $n$ gets larger the boomerang RG flows build up an increasingly larger intermediate scaling regime that is determined by the $AdS_2\times\mathbb{R}^3$ solution, as also displayed in the radial behaviour of the scalar field $\gamma$ in figure \ref{othercase}.

For the values of $\Gamma/k^{3/2}$ where there is not a unique solution, the physical RG flow solution is the one that has the
smallest energy. In figure \ref{othercase} we have plotted $T^{tt}$ for the boomerang flows and we find that for a given value of $\Gamma/k^{3/2}$
the energetically preferred solution is given by the smallest value of $n$.
In particular, the amount of build up of an intermediate scaling regime determined by the $AdS_2\times\mathbb{R}^3$ solution is frustrated for energetic reasons.

We have constructed some finite temperature black hole solutions, but the full phase diagram is rather involved due to the presence
of multiple branches of solutions in the region of $\Gamma/k^{3/2}$ where there is non-uniqueness of the boomerang RG flows. While
we leave a full analysis to future work, we note that we have constructed some black hole solutions for values of $\Gamma/k^{3/2}$
significantly larger than $\bar\Gamma$ and we find that as $T\to 0$ the ground states have $s(T)\to 0$, not 
as a power law, and
they are again thermal insulators. Furthermore, we also find that they satisfy the diffusion-butterfly velocity relation given in \eqref{dvbee} with again, remarkably, $E(T)\to 0.5$ as $T\to 0$.


\providecommand{\href}[2]{#2}\begingroup\raggedright\endgroup

\end{document}